\documentclass{aa}  
\usepackage{comment}
\usepackage{natbib}
\usepackage{multirow}
\usepackage{booktabs}
\usepackage{graphicx}
\graphicspath{{Fig/}}
\usepackage[table]{xcolor}
\usepackage{menukeys}
\usepackage{array, multirow, bigdelim, makecell, booktabs} 

\DeclareUnicodeCharacter{2212}{-}
\usepackage{txfonts}

\newcommand{\kms}{km~s$^{-1}$}

\makeatletter
\renewcommand*\aa@pageof{, page \thepage{} of \pageref*{LastPage}}
\makeatother
\usepackage[colorlinks=true, allcolors=blue]{hyperref}

\begin{document}

\title{The CH radical at radio wavelengths: Revisiting emission in the 3.3 GHz ground state lines}
    \author{Arshia M. Jacob\thanks{Member of the International Max Planck Research School (IMPRS) for Astronomy and Astrophysics at the Universities of Bonn and Cologne}
          \and
          Karl M. Menten
          \and 
          Helmut Wiesemeyer
          \and 
          Gisela N. Ortiz-Le\'{o}n
          }

   \institute{Max-Planck-Institut f\"{u}r Radioastronomie, Auf dem H\"{u}gel 69, 53121 Bonn, Germany
    \email{ajacob@mpifr-bonn.mpg.de}}

   \date{Received 25 January, 2021/ Accepted April 9, 2021}
   \titlerunning{The CH radical at radio wavelengths}
   \authorrunning{A. Jacob et al.}
 
  \abstract{
  The intensities of the three widely observed radio-wavelength hyperfine structure (HFS) lines between the $\Lambda$-doublet components of the rotational ground state of CH are inconsistent with local thermodynamic equilibrium (LTE) and indicate ubiquitous population inversion. While this can be qualitatively understood assuming a pumping cycle that involves collisional excitation processes, the relative intensities of the lines and in particular the dominance of the lowest frequency satellite line has not been well understood. This has limited the use of CH radio emission as a tracer of the molecular interstellar medium.
  }
  {
  We aim to investigate the nature of the (generally) weak CH ground state masers by employing synergies between the ground state HFS transitions themselves and with the far-infrared lines, near 149~$\mu$m (2~THz), that connect these levels to an also HFS split rotationally excited level.
  } 
  {
  We present the first interferometric observations, with the \textit{Karl G. Jansky} Very Large Array, of the CH 9~cm ground state HFS transitions at 3.264~GHz, 3.335~GHz, and 3.349~GHz toward the four high mass star-forming regions (SFRs) Sgr~B2~(M), G34.26+0.15, W49~(N), and W51.  We combine this data set with our high spectral resolution observations of the $N,J\!=\!2,3/2\!\rightarrow\!1,1/2$ transitions of CH near 149~$\mu$m observed toward the same sources made with the upGREAT receiver on SOFIA, which share a common lower energy levels with the HFS transitions within the rotational ground state. 
  }
  {
  Toward all four sources, we observe the 3.264~GHz lower satellite line in enhanced emission with its relative intensity higher than its expected value at LTE by a factor between 4 and 20. Employing recently calculated collisional rate coefficients, we perform statistical equilibrium calculations with the non-LTE radiative transfer code MOLPOP-CEP in order to model the excitation conditions traced by the ground state HFS lines of CH and to infer the physical conditions in the emitting regions. The models account for effects of far-infrared line overlap with additional constraints provided by reliable column densities of CH estimated from the 149~$\mu$m lines.
  }
  {
  The derived gas densities indicate that the CH radio emission lines (and the far-infrared absorption) arise from the diffuse and translucent outer regions of the SFRs' envelopes as well as in such clouds located along the lines of sight. We infer temperatures ranging from 50 to 125~K. These elevated temperatures, together with astrochemical considerations, may indicate that CH is formed in material heated by the dissipation of interstellar turbulence, which has been invoked for other molecules. The excitation conditions we derive reproduce the observed level inversion in all three of the ground state HFS lines of CH over a wide range of gas densities with an excitation temperature of ${\sim\!-0.3~}$K, consistent with previous theoretical predictions. }

\keywords{ISM: molecules -- ISM: abundances -- ISM: clouds -- ISM: lines and bands -- methods: numerical -- radiative transfer}

   \maketitle

\section{Introduction}\label{sec:intro}
The 4300.3~\AA\ electronic transition of the methylidyne radical, CH, was one of the first three molecular lines detected in the interstellar medium \citep[ISM;][]{dunham1937interstellar, swings1937, McKellar1940}. Since then, CH has been extensively observed in a wide range of wavelength regimes, from the radio at 9~cm (3.3 GHz) \citep{rydbeck1973}, over the sub-millimetre (sub-mm) and far-infrared (FIR) ranges to the far-ultraviolet (FUV) regime at 1369.13~\AA\ \citep{watson2001assignment}. The CH radical has also been detected in a variety of environments, hosting different physical and excitation conditions from diffuse, and translucent interstellar clouds to dark clouds, outflows, and shocks at the edges of dense molecular clouds, and even in external galaxies \citep[for example,][]{Lang1978, Whiteoak1980, Mattila1986, Sandell1988, Magnani1992, Magnani1993}. 

As the simplest carbyne, CH is an important intermediary in the gas phase chemistry of carbon bearing molecules, initiating the formation of larger and more complex species in the ISM. In addition, studies using high resolution optical spectroscopy were able to reveal a tight correlation between the derived column densities of CH, and those of H$_{2}$ (or the visual extinction) in diffuse and translucent clouds \citep{federman1982measurements, sheffer2008, Weselak2019}. However, such optical absorption studies are limited to nearby clouds (a few kpc) as they require visually bright ($V\!<\!10$~mag) background stars. Fortunately, the advent of space- and air-borne telescopes like Herschel and SOFIA, respectively, have not only renewed interests in CH and other light hydrides, but have also extended their studies over Galactic scales. Measurements of absorption in the generally optically thin rotational transitions of CH at 532/536~GHz (560~$\mu$m) \citep{gerin2010interstellar} and 2006/2010~GHz (149~$\mu$m) \citep{wiesemeyer2018unveiling,jacob2019fingerprinting} against the continuum emission from distant star-forming regions (SFRs) in spiral arms and the Galactic centre region, yielding column densities, have further emphasised its use as a tracer for H$_{2}$ in diffuse and translucent clouds.\\

In the early days of molecular radio astronomy, \citet{rydbeck1973} first detected the hyperfine structure (HFS) lines between the $\Lambda$-doublet levels of the rotational ground state of CH at 3.3~GHz (9~cm). At frequencies near 3.264~GHz, 3.335~GHz, and 3.349~GHz (see Table~\ref{tab:spec_properties}), they were always detected in (generally quite weak) emission. Extensive single dish surveys across the Galaxy found the CH radio emission to be very widespread in the general molecular ISM in regions ranging from quiescent dark clouds to the environment of H{\small II} regions \citep[for example,][]{Zuckerman1975, Rydbeck1976, Genzel1979}.

\begin{table}
    \centering
    \caption{Spectroscopic properties of the CH ground state HFS transitions. The columns are (from left to right): the transition as described by the hyperfine quantum number ($F$), the frequency of the transition, the Einstein A coefficient and the relative line intensities at LTE.}
    \begin{tabular}{cccc}
    \hline \hline 
       Transition  &   Frequency  &  $A_{\text{E}}$ & Relative \\
       $F^{\prime} - F^{\prime\prime}$ & [MHz] & $\times10^{-10}$~[s$^{-1}$] & Intensity\\ 
         \hline 
        $0^{-}-1^{+}$ & 3263.793447 & 2.876 & 1.0 \\
        $1^{-}-1^{+}$ & 3335.479356 & 2.045 & 2.0 \\
        $1^{-}-0^{+}$ & 3349.192556 & 1.036 & 1.0 \\
         \hline 
    \end{tabular}
    \tablefoot{The frequencies were measured by \citet{Truppe2014} with uncertainties of 3~Hz.}
    \label{tab:spec_properties}
\end{table}

That always emission was observed, even toward continuum sources, and that the relative populations of the three HFS lines often deviate from their expected values at local thermodynamic equilibrium (LTE), which are $I_{3.264}:I_{3.335}:I_{3.349} = 1:2:1$ (see Table~\ref{tab:spec_properties}), suggested that the populations of the CH ground state $\Lambda$-doublet HFS levels must be inverted. The observed ubiquitous inversion of these CH transitions across sources for which one might assume varying physical conditions suggest that there must exist a general pumping mechanism that preferentially populates the upper HFS levels of the ground state $\Lambda$-doublet independent of the prevailing physical conditions within a region. The level inversion, leading to weak maser action in the ground state HFS line of CH was initially thought to be excited through collisional processes (involving collisions with atomic or molecular hydrogen) to the first rotational level \citep{Bertojo1976, Elitzur1977}. However, \citet{Bujarrabal1984} have shown that collisions alone cannot explain the observed excitation anomalies. In particular, the $F^{\prime}-F^{\prime\prime}=0^{-}-1^{+}$ satellite line at 3.264~GHz is seen in enhanced emission while the $F^{\prime}-F^{\prime\prime}=1^{-}-1^{+}$ main line may sometimes, but very rarely, appear in absorption against the strong continuum of background sources. Excitation by collisions alone cannot be solely responsible for the observed `enhancement' in the relative intensity of the lower satellite line, and necessitates the inclusion of radiative processes in order to explain the observed line strengths of the CH ground state transitions in SFRs.

\begin{figure}
    \includegraphics[width=0.5\textwidth]{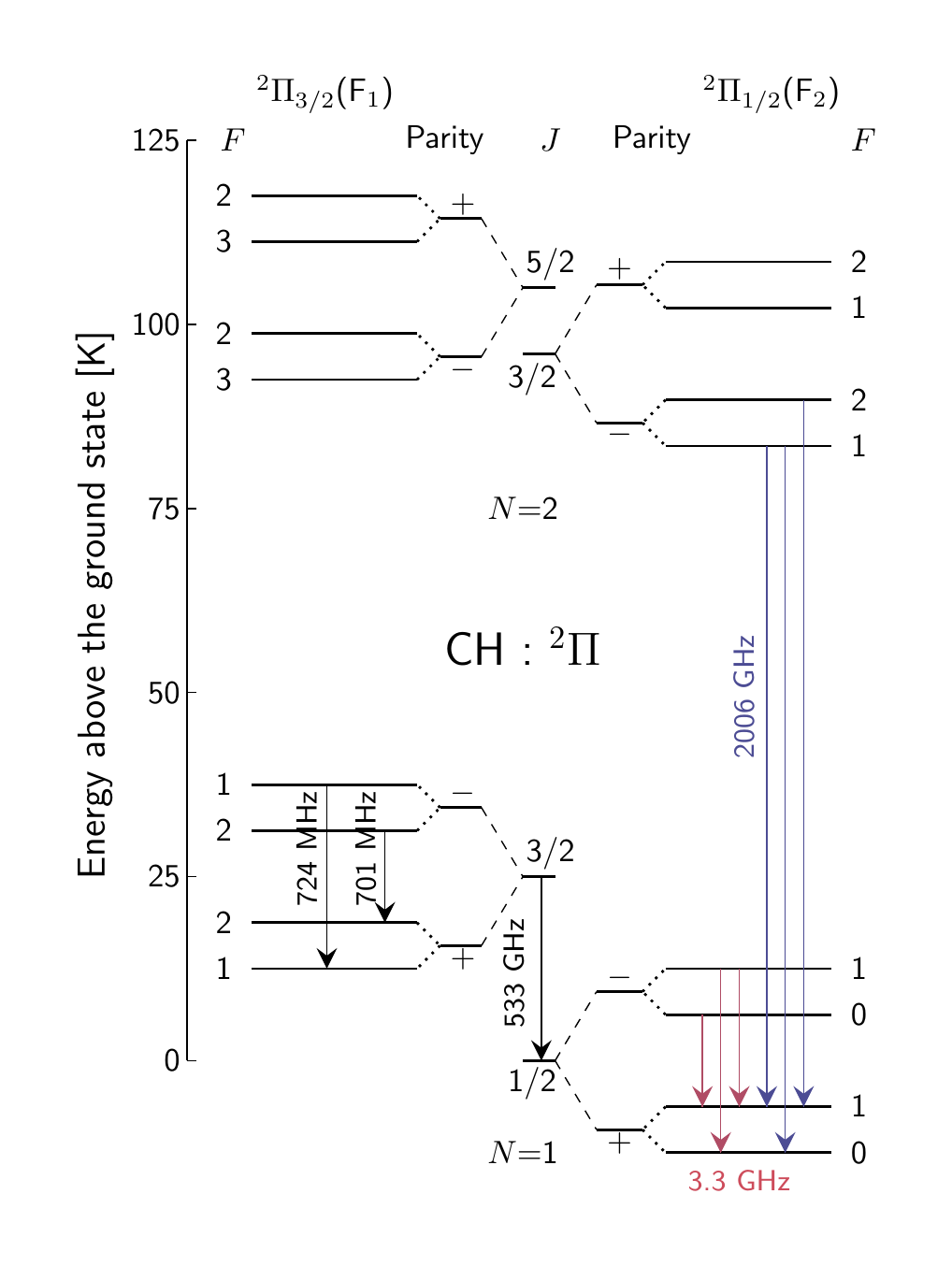}
    \caption{Lowest rotational energy levels of CH. The relevant transitions are labelled with their transitions marked using arrows, with the HFS splitting of the ground state energy level of CH as well as the rotational transition from the $N,J\!=\!2,3/2\!\rightarrow\!1,1/2$ highlighted in magenta and blue, respectively. Note that the $\Lambda$-doublet and HFS splitting level separations are not drawn to scale.}
    \label{fig:energy_level}
\end{figure}

In this work, we aim to investigate the excitation mechanism causing the ubiquitous weak masering effects in the CH ground state lines. Our analysis is aided by recently calculated collisional rate coefficients determined for inelastic collisions of CH with H, H$_{2}$, and He by \citet{Dagdigian2018} and \citet{Marinakis2019}. Moreover, the use of accurate column density measurements determined from the high angular resolution observations of the $N,J\!=\!2,3/2\rightarrow\!1,1/2$ FIR transitions of CH near 149~$\mu$m (2006~GHz) observed using the upGREAT instrument \citep{risacher2016upgreat}  on board the Stratospheric Observatory for Infrared Astronomy \citep[SOFIA;][]{young2012early}, whose HFS components have a common lower energy level with the CH radio ground state lines (see Fig.~\ref{fig:energy_level}), will provide new constraints on our non-LTE models. In principle, such a comparison between the 2006~GHz transitions of CH observed using the 91.4~cm telescope onboard the Kuiper Airborne Observatory (KAO) and the 3.264~GHz ground state line, was previously carried out by \citet{Stacey1987} towards Sgr~B2~(M). While the analysis presented by these authors attributes the level inversion to excitation effects and subsequently estimates the excitation temperature, the accuracy of their estimations is limited by the coarse spectral resolution of their FIR observations. The higher spatial and spectral resolution of our data allows us to estimate the column densities of CH and broadly assign contributions as arising from different spiral-arm and inter-arm regions for any given sight line. 

We present here, the first interferometric observations of CH using the NRAO\footnote{The National Radio Astronomy Observatory (NRAO) is operated by Associated Universities Inc., under a collaborative agreement with
the US National Science Foundation.} \emph{Karl G. Jansky} Very Large Array (VLA) in New Mexico toward four well known high-mass SFRs, namely, Sgr~B2~(M), G34.26+0.15, W49~(N), and W51. In Sect.~\ref{sec:excitation_mech} we detail the excitation mechanism of the CH ground state and describe the observational setup in Sect.~\ref{sec:observations}. We present the resulting line profiles, and introduce our non-LTE models in Sect.~\ref{sec:results} and discuss the subsequently obtained physical and excitation conditions in Sect.~\ref{sec:discussion} and finally in Sect.~\ref{sec:conclusions} we summarise the main conclusions derived from this work. 

\section{Excitation mechanism of the CH  ground state}
\label{sec:excitation_mech}
The ground electronic state of the CH radical, with an unpaired $\pi$ electron in its valence shell, exists in a $^{2}\Pi$ state. Following Hund's case b coupling, the spin-orbit interaction ($\boldsymbol{L.S}$) splits each principal quantum level, $N$, into two spin-orbital manifolds -- ${}^{2}\Pi_{1/2}(F_{2})$ and ${}^{2}\Pi_{3/2}(F_{1})$, respectively, as displayed in Fig.~\ref{fig:energy_level}. For CH, unlike in the case for OH, the ${}^{2}\Pi_{3/2}$ ladder is above the ${}^{2}\Pi_{1/2}$ ladder, that is, the absolute ground state is in the ${}^{2}\Pi_{1/2}, J=1/2$ level. The spatial orientation of the electron's orbital momentum axis relative to the axis of the molecule's rotation, splits the rotational levels, $J$, into $\Lambda$-doublet states. The $\Lambda$-doublet levels are distinguished based on their parity denoted by either + or $-$. Interactions between the total angular momentum and the nuclear spin of the hydrogen atom ($I_{\rm H} = 1/2$) further split each $\Lambda$-doublet level into two HFS levels with hyperfine quantum numbers, $F$. In general, for the ${}^{2}\Pi_{1/2}(F_{2})$ ladder, the energies of the HFS levels are ordered with increasing values of $F$ while for the ${}^{2}\Pi_{3/2}(F_{1})$ ladder the opposite holds for all HFS levels, except for the lowest. See \citet{Truppe2014} for more details on the electronic structure of the CH ground state.\\

As briefly discussed in Sect.~\ref{sec:intro}, the widely observed level inversion in CH can be qualitatively understood by invoking a simple pumping cycle. The pumping cycle involves the collisional excitation of CH (via collisions with atomic hydrogen, H$_{2}$, He and/or electrons) to the first rotational state, $^{2}\Pi_{3/2}(F_{1}),J\!=\!3/2$, followed by radiative decay back to the ground state \citep{Bertojo1976, Elitzur1977}. Moreover, in order for collisions to produce a level inversion, the excitation cross-sections of the lower level of the $^{2}\Pi_{3/2}(F_{1}), J\!=\!3/2$ $\Lambda$-doublet level must be larger than those of its upper level such that the preferential population of this level will create a level inversion after it cascades to the ground state. This difference in the collisional cross-sections within pairs of $\Lambda$-doublets is referred to as `parity discrimination' \citep[see,][for a detailed description]{Dixon1979}. In terms of parity discrimination, the pumping cycle works such that collisions between the + parity components of the $\Lambda$-doublet levels, corresponding to the $^{2}\Pi_{1/2}(F_{2}), J\!=\!1/2$ and $^{2}\Pi_{3/2}(F_{1}), J\!=\!3/2$ levels, are more probable than collisions between their $-$ components. The radiative cascade that subsequently follows ($+\!\rightarrow\!-$ and $-\!\rightarrow\!+$) will therefore tend to overpopulate both HFS levels of the $-$ parity component of the ground state $\Lambda$-doublet relative to the $+$ parity components HFS levels.

Theoretical considerations by \citet{Bertojo1976} and \citet{Dixon1979, Dixon1979inelastic} have shown that the effects of parity discrimination-induced level inversion vary depending on the collisional partner considered. Both studies were in agreement that, with a smaller reduced mass, collisions between CH and atomic hydrogen effectively invert the ground state HFS levels, but they disagreed on the role played by collisions with H$_{2}$. While \citet{Bertojo1976} suggest that collisions of CH with H$_{2}$ will tend to follow a similar parity discrimination as those for the case of H, \citet{Dixon1979inelastic} suggest an inverse propensity for collisions with H$_{2}$, where the collision between the $-\!\rightarrow\!-$ levels dominate. The latter contradicts observations of the ground state transitions of CH toward SFRs which show anomalous excitation effects at velocities corresponding to the envelopes of molecular clouds which are mainly comprised of molecular material. Taking all things into consideration the models we discuss in Sect.~\ref{subsec:nlte_models} do not specifically take into account parity discrimination effects but rather weigh contributions from different collision partners based on the molecular fraction ($f_{\text{H}_{2}}$) of the clouds considered.

Following selection rules, the ground state level of CH displays three transitions as the result of HFS splitting, at 3.264~GHz, 3.335~GHz, and 3.349~GHz, respectively (the spectroscopic properties of which are summarised in Table~\ref{tab:spec_properties}). Because it has twice the theoretical (LTE) intensity of the two other lines, the 3.335~GHz transition corresponding to the ${F^{\prime}-F^{\prime\prime}\!=\!1^{-}-1^{+}}$ is referred to as the main HFS line, while the other two transitions corresponding to ${F^{\prime}-F^{\prime\prime}\!=\!0^{-}-1^{+}}$ and ${F^{\prime}-F^{\prime\prime}\!=\!1^{-}-0^{+}}$ at 3.264~GHz, and 3.349~GHz are known as the lower, and upper satellite lines, respectively. While the collisional pumping model discussed above describes the framework for the level inversion observed in the ground state of CH, it does not account for the relative line strengths between its different HFS components, more specifically that of the enhanced intensity observed in the lower satellite line, which generally is the strongest of the three.  As discussed in \citet{Zuckerman1975} and \citet{Bujarrabal1984}, the HFS lines' relative intensities are influenced by effects of line overlap of transitions to higher rotational levels. This can occur either through thermal broadening of the lines or via bulk velocity gradients in the gas, both of which are capable of bringing different sets of transitions into resonance with one another. With a typical line width between 2~km~s$^{-1}$ (for line-of-sight (LOS) components) and $\geq~$6~km~s$^{-1}$ (at the envelope of the molecular cloud), there maybe partial or total line overlap from the HFS lines from the $+$ parity component of the $\Lambda$-doublet levels of the first rotationally excited states of both orbital manifolds of CH at $^{2}\Pi_{3/2}(F_{1}), J=3/2$ and $^{2}\Pi_{1/2}(F_{2}), J=3/2$. Therefore, it is essential to take into account effects of line overlap as they significantly alter the amount of radiative trapping between the two $\Lambda$-doublet levels of the ground state.

\section{Observations}\label{sec:observations}

\begin{table*}
\centering
  \caption{Summary of source parameters. The columns are (from left to right): the source designation, the equatorial source coordinates, the heliocentric distances, distance reference, systemic velocity of the source, FWHM of the synthesised beams ($\theta_{\rm maj} \times \theta_{\rm min}$), position angles (P.A.), and FWHM of the restored circular beams ($\theta_{\rm B}$) and the root-mean-square (rms) noise levels of the line and continuum maps, respectively.}
    \begin{tabular}{lrrrrrccccc}
    \hline \hline 
         \multicolumn{1}{c}{Source} & \multicolumn{2}{c}{Coordinates (J2000)}  & $d$ & Ref & $\upsilon_{\text{LSR}}$ & \multicolumn{1}{c}{$\theta_{\rm maj} \times \theta_{\rm min}$} & P.A. & \multicolumn{1}{c}{$\theta_{\rm B}$} & rms$_{\rm line}$ & rms$_{\rm cont}$\\ 
           &   $\alpha$~[hh:mm:ss] & $\delta$~[dd:mm:ss]  
           & [kpc] & & [km~s$^{-1}$] & \multicolumn{1}{c}{[$^{\prime\prime}\times{}^{\prime\prime}$]} & [$^{\circ}$]&  \multicolumn{1}{c}{[$^{\prime\prime}$]} & [mJy/beam] & [mJy/beam] \\
         \hline  
         Sgr~B2~(M) & 17:47:20.50 & $-$28:23:06.00 & 8.3 & [1] & 64.0 & $35.3 \times 15.0$ & 169.7 & 23.0 & 2.5 & 2.7 \\
         G34.26+0.15 & 18:53:18.70 & +01:14:58.00 & 1.6 & [2]  & 59.0 & $22.3 \times 18.8$& $-$26.7& 20.5 & 2.0  & 1.1\\
         W49~(N) & 19:10:13.20 & +09:06:11.88 & 11.4 & [3] & 11.8  & $20.5 \times 18.6$ & $-32.2$ & 19.5 & 2.6 & 4.1     \\
         W51~E & 19:23:43.90 & +14:30:31.00 & 5.4 & [4] & 62.0 & $33.2 \times 23.2$ & $-$48.5& 27.7 & 3.2 & 3.1 \\
         \hline 
         \end{tabular}
  \tablebib{For the heliocentric distances:[1]~\citet{Reid2019};[2]~\citet{zhang2009trigonometric};[3]~\citet{zhang2013parallaxes};[4]~\citet{sato2010}}
 
    \label{tab:source_parameters}
\end{table*}

Observations of the HFS lines between the $\Lambda$-doublet of the ground state level of CH near 3.3~GHz were carried out on 2017 March 3, 5, 16, and 23 using the radio `S-band' receivers of the VLA in the D-configuration (project id: 17A-214) for a total observing time of 1.3~hours per source. A total of eight well known SFRs as well as an active galactic nucleus (AGN) were observed across the four observational epochs mentioned above. However in this first paper we present the results and analysis toward only four of these sources, namely, Sgr~B2~(M), G34.26+0.15, W49~(N), and W51. This subset of sources, all of which are characterised by strong mm and FIR continuum emission, was selected in particular, not only because the physical conditions toward them have been extensively studied but, importantly, also because complementary observations of the FIR $N,J = 2,3/2 \rightarrow 1,1/2$ HFS transitions of CH at 2006~GHz toward them had been made with the upGREAT receiver on board SOFIA. The main properties of the sources discussed in our study are summarised in Table~\ref{tab:source_parameters}. 

The VLA WIDAR correlator was configured to allow three spectral setups consisting of a 4~MHz wide sub-band or 2048 channels, centred at the frequency of each of the three CH lines, the spectroscopic parameters of which are summarised in Table~\ref{tab:spec_properties}. This setup corresponds to a native spatial resolution of 1.953~kHz (corresponding to 0.18 km~s$^{-1}$). We also carry out broad band continuum observations over a total bandwidth of 2~GHz between 1.98 and 3.98~GHz, in 16 spectral windows, each of which are 128~MHz wide. The quasar 3C286 was used as both the bandpass as well as the flux calibrator, while J1751$-$2524, J1830+0619, J1924+1540, and J1925+2106 were used as phase calibrators for the different epochs in which Sgr~B2~(M), G34.26+0.15, W49~(N), and W51 were observed. In addition to the three CH spectral setups, there were eight spectral window setups used to observe hydrogen radio recombination lines (HRRLs) with principal quantum numbers, $n$, between 123 and 130. However, since this work mainly concentrates on the CH ground state transitions, the HRRLs will not be discussed here. 

The data were calibrated and imaged using the Common Astronomy Software Applications (CASA) package with a modified version of the VLA pipeline\footnote{\url{https://science.nrao.edu/facilities/vla/data-processing/pipeline}}. Then, the calibrated visibilities were CLEANed with a pixel size of $3^{\prime\prime}$ and an image size of $15^{\prime}\times 15^{\prime}$. Continuum subtraction was performed to the visibility data using the task {\tt uvcontsub}, for which we exclude channels with line emission or absorption. To allow a meaningful comparison with the SOFIA data, the VLA images were restored with circular beams of full-width at half-maximum (FWHM) $\theta_{\rm B}$ with areas identical to those of the elliptical beams synthesised by CASA, that is, $\theta_{\rm B} = \sqrt{\theta_{\rm maj} \times \theta_{\rm min}}$, where $\theta_{\rm maj}$ and  $\theta_{\rm min}$ are, respectively, the major and minor FWHM axes of the synthesised beam. Values for $\theta_{\rm B}$, $\theta_{\rm maj}$ and 
$\theta_{\rm min}$ are given in Table~\ref{tab:source_parameters} following the sources' positions and other attributes. For the analysis that follows, the intensity scales were converted from specific brightness, $S$, to brightness temperature, $T_{\rm B}$, using the Rayleigh-Jeans relation,
\begin{equation}
T_{\rm B} = \frac{S c^2}{2\nu^2~k_{\rm B}~\Omega} = \frac{1.222\times10^{6}~{\rm K}}{\nu^{2}~({\rm GHz)}~\theta_{\rm B}^2(^{\prime\prime})} S ~({\rm Jy~beam}^{-1})\, ,
\end{equation}
where $k_{\rm B}$ is the Boltzmann constant, $c$ the speed of light, $\nu$ the frequency and $\Omega$ the beam solid angle. The second expression allows for convenient units to obtain $T_{\rm B}$ in K, from $S$ measured in Jy~beam$^{-1}$, observed at a frequency $\nu$ (in GHz) with a circular beam with a FWHM beam size of $\theta_{\rm B}$ in arc seconds. For $\nu = 3.3~$GHz and our average $\theta_{\rm B} = 23^{\prime\prime}$, $T_{\rm B}/S$ = 212~K/Jy.

The spectra of the 3.3~GHz CH lines were extracted from a region encompassing the same area as that by the FWHM beam width of the central pixel of the upGREAT/SOFIA (13.5$^{\prime\prime}$) and centred at the same positions toward which the 2006~GHz observations were carried out. These pointing positions coincide with the centres of the map, the corresponding positions of which, toward the different sources are tabulated in Table~\ref{tab:source_parameters}. The spectra were further post-processed and analysed using Python packages Astropy \citep{Astropy2013, Astropy2018} and APLpy \citep{AplPy2012}. The spectra of the CH transitions near 2006~GHz presented in the following sections do not represent the actually observed spectra but rather the line profiles resulting from the deconvolution of its HFS splitting. The HFS is deconvolved following the Wiener filter algorithm discussed in \citet{jacob2019fingerprinting} with the local standard of rest (LSR) velocity scale set by the strongest HFS component of the 2006~GHz transition.

\section{Results}\label{sec:results}
\begin{figure*} 
    \includegraphics[width=0.48\textwidth]{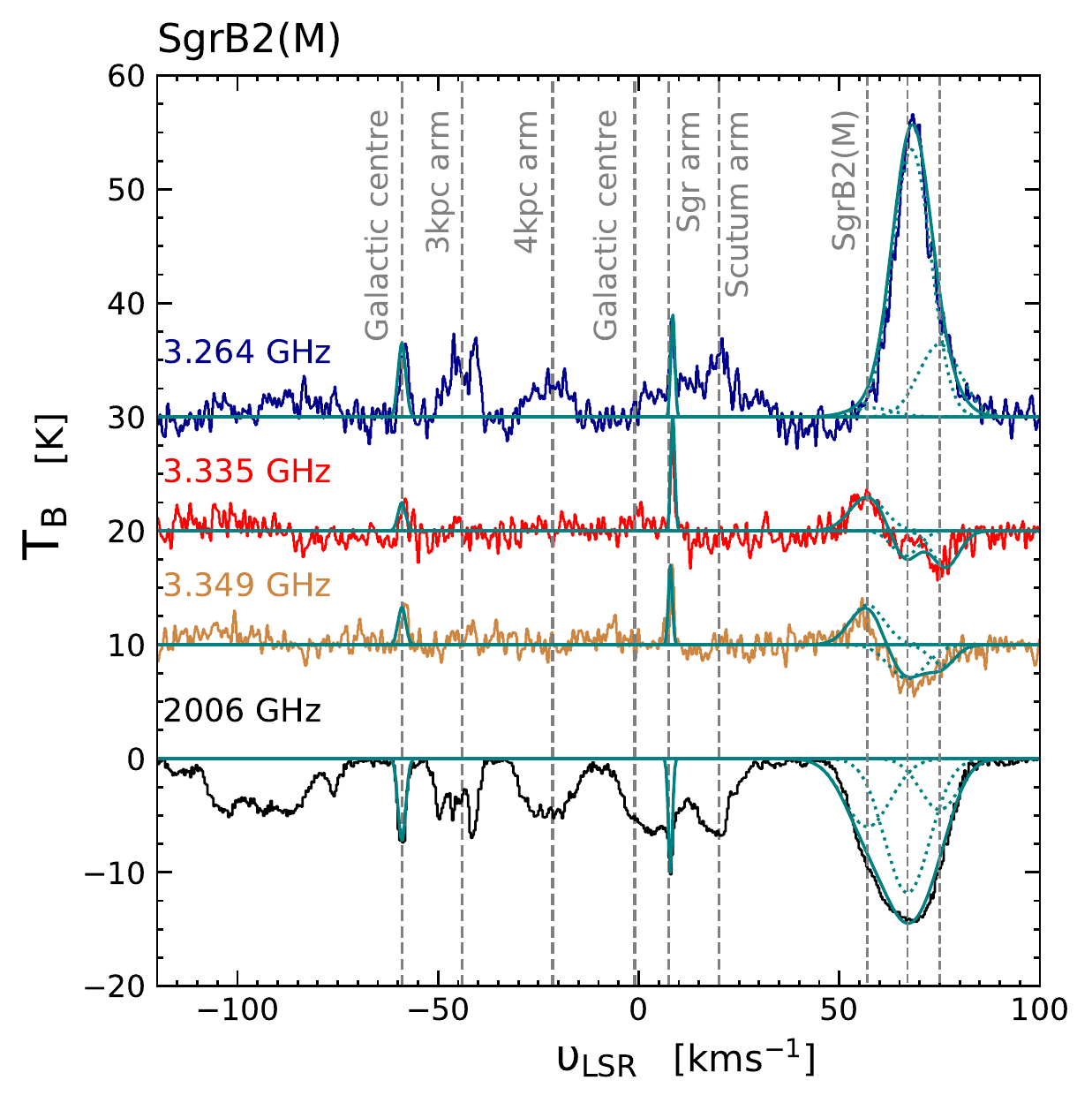}\quad
    \includegraphics[width=0.475\textwidth]{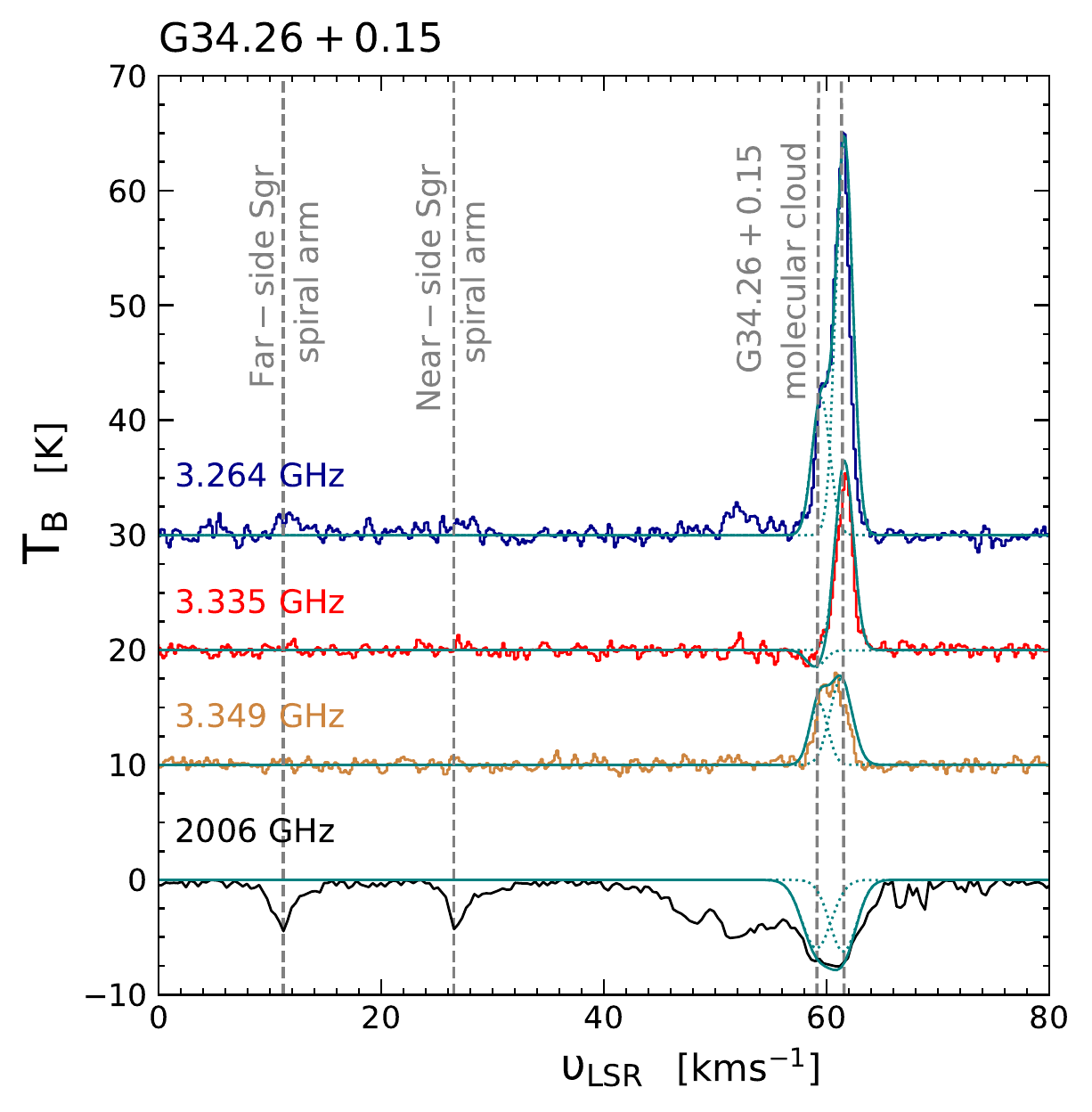}
    \caption{Spectra of the ground state HFS transitions of CH at 3.264~GHz (in blue), 3.335~GHz (in red), and 3.349~GHz (in dark orange) and the $N,J = 2,3/2 \rightarrow 1,1/2$ HFS deconvolved CH spectrum near 2006~GHz (in black) toward Sgr~B2~(M) (left) and G34.26+0.15 (right). The LSR velocity scale for the 2006~GHz transition is set by the strongest HFS component. The spectra of the ground state lines are offset from the baseline by 30, 20, and 10~K. The dashed grey lines mark the main spiral- and inter-arm components crossing the LOS toward each sight line, as discussed in the text. Gaussian fits to LOS velocity features that are common to all four transitions are displayed by the solid teal curve and the dotted teal curves indicate the different components observed within the envelope of the Sgr~B2~(M), and G34.26+0.15 molecular clouds, respectively.}
    \label{fig:SgrB2_G34P26_spec}
\end{figure*}

\begin{figure*}
    \centering
    \includegraphics[width=0.47\textwidth]{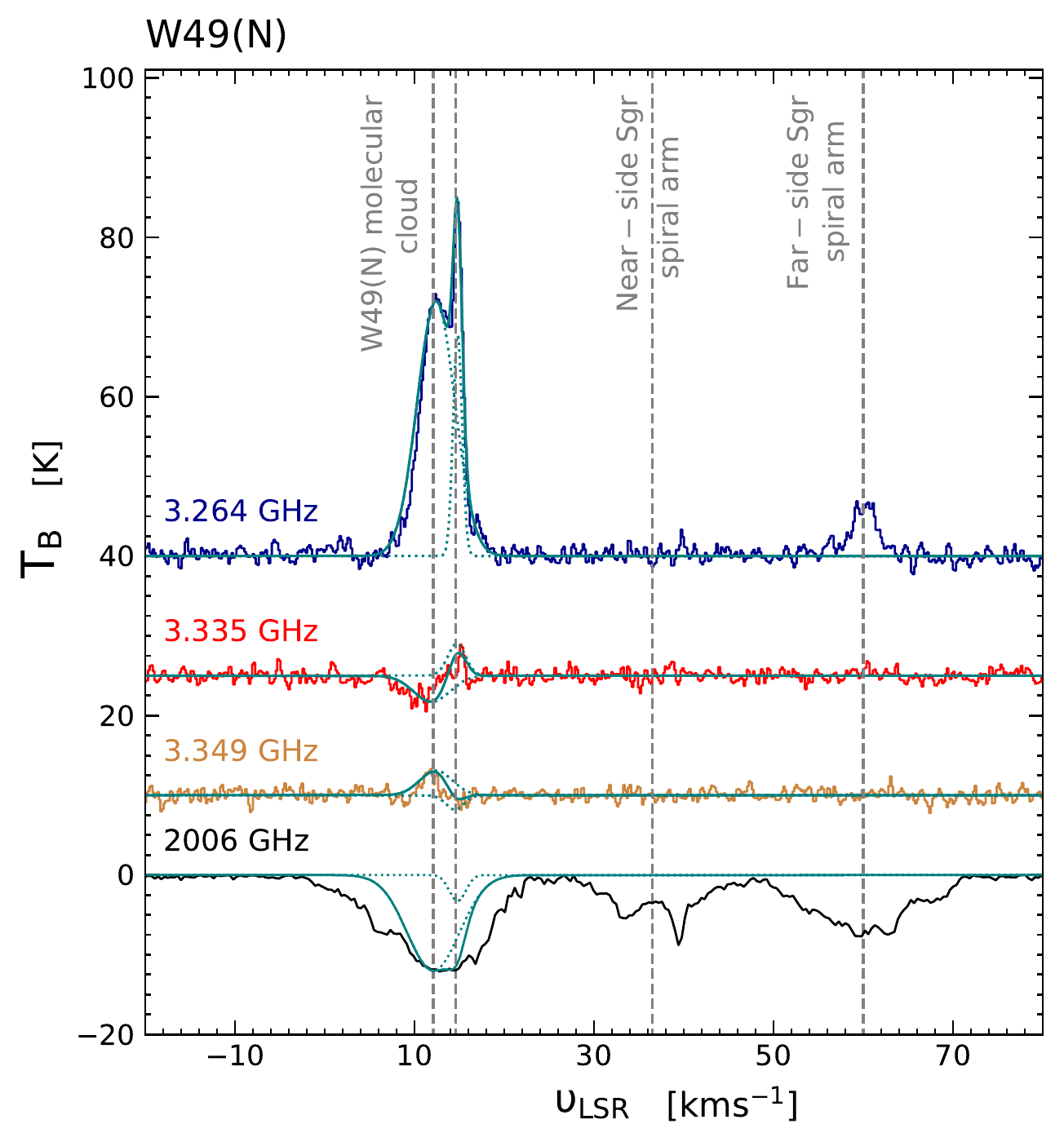}\quad
    \includegraphics[width=0.482\textwidth]{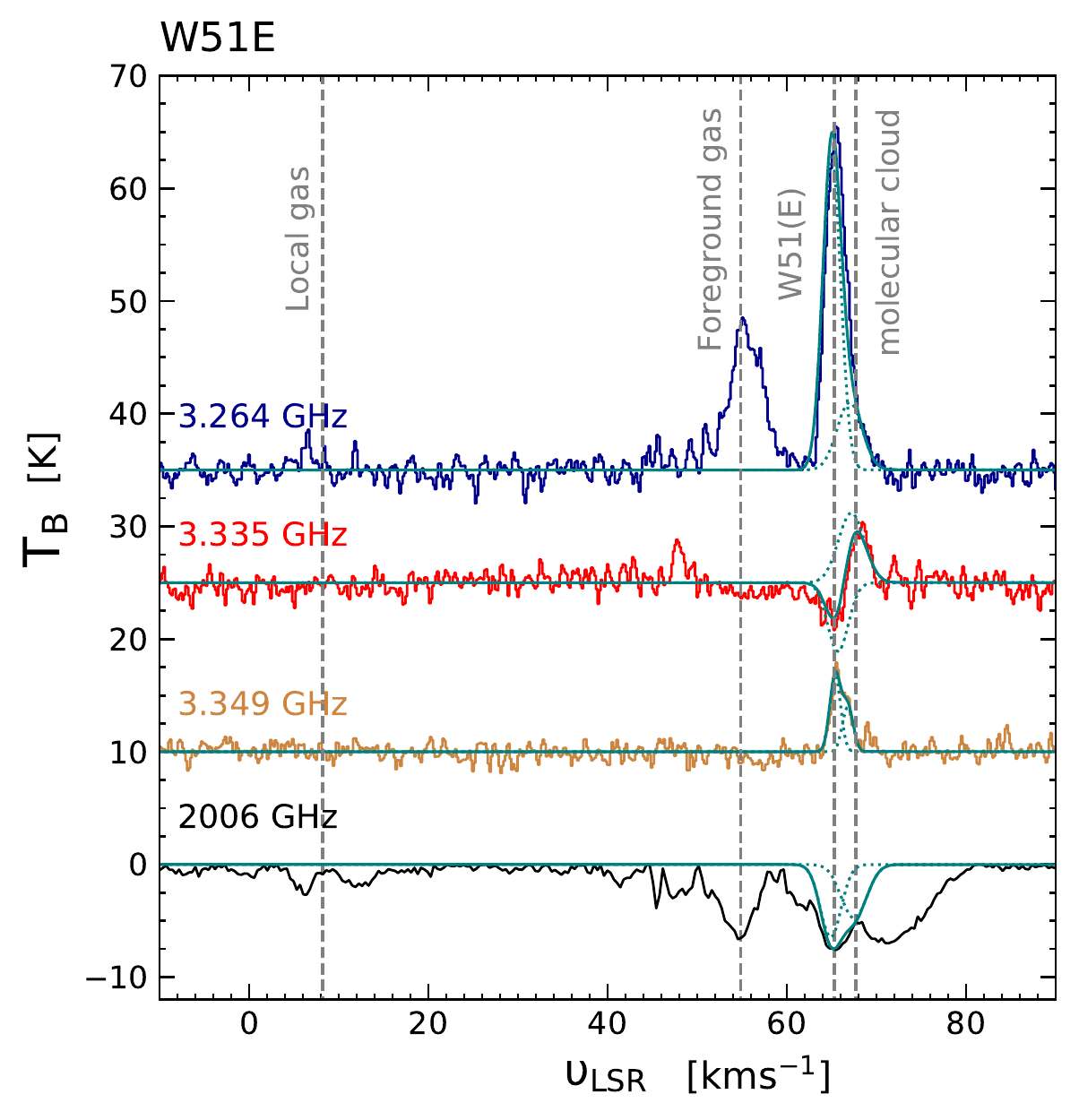}
    \caption{Same as Fig.~\ref{fig:SgrB2_G34P26_spec} but toward W49~(N) (left), and W51~E (right). Here, the spectra of the ground state lines are offset from the baseline by (40, 35, 10, and $-20$~K) and (35, 25, 10, and $-12$~K), for W49~(N) and W51~E, respectively. }
    \label{fig:W49N_W51_spec}
\end{figure*}

\begin{figure*}
    \centering
    \includegraphics[width=1\textwidth]{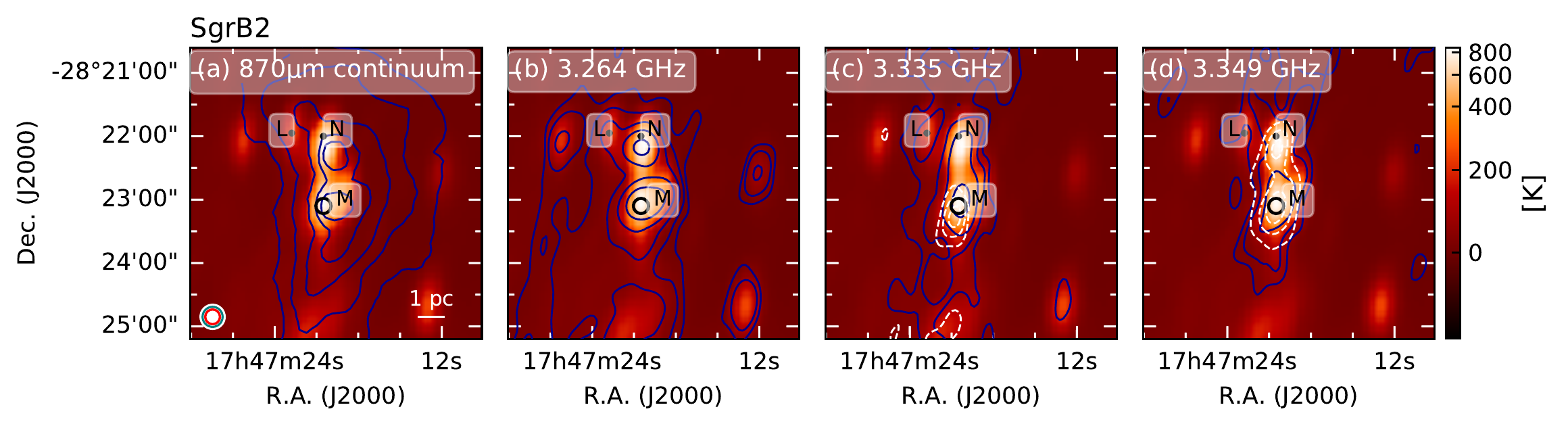}
    \caption{Overview of the 3.3~GHz CH line observations toward Sgr~B2. Panel (a) displays the ATLASGAL 870~$\mu$m emission at contour levels starting from $5\times, 10\times, 20\times, 40\times$ and $80\times 1 \sigma$ where $1\sigma = 0.9~$Jy/beam overlaid atop the 3.3~GHz continuum emission along with the synthesised beam of the VLA observations (filled white circle) as well as the central beam of upGREAT/SOFIA (red) and the LABOCA/APEX beam (teal), for comparison. The position from which the spectra are extracted is marked by the beam in black and labelled M, we also mark the positions of Sgr~B2~(N) and (L). Panels (b), (c), and (d) display the integrated intensity contours of the 3.264~GHz, 3.335~GHz, and 3.349~GHz CH lines, respectively, overlaid on top of the 3.3~GHz continuum emission. The intensities are integrated over a velocity range between 48 and 83~km~s$^{-1}$ except for that of the 3.335~GHz and 3.349~GHz CH maps, for which the emission components are integrated between 48 and 62~km~s$^{-1}$ while the absorption between between 62 and 84~km~s$^{-1}$. The contour levels mark $5\times, 10\times, 20\times, 40\times,$ and 80$\times 1 \sigma$ where $1\sigma = 10.6, 2.65,$ and 2.65~K~km~s$^{-1}$ for the 3.264~GHz, 3.335~GHz, and 3.349~GHz CH emission components, respectively (solid dark blue curves) and the $3\times, 4\times,$ and 5$\times1\sigma$ levels where $1\sigma=2.12$~K~km~s$^{-1}$ for the 3.335~GHz and 3.349~GHz CH absorption components (dashed white curves).  }
    \label{fig:SgrB2_map}
\end{figure*}

\begin{figure*}
    \centering
    \includegraphics[width=1\textwidth]{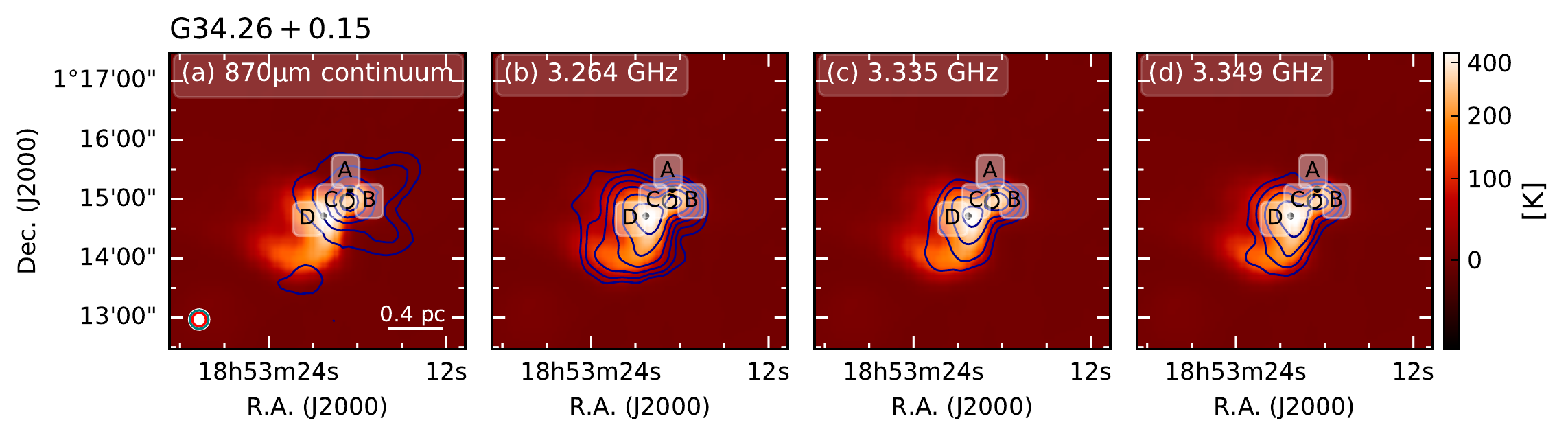}
    \caption{Overview of the 3.3~GHz CH line observations toward G34.26+0.15. Panel (a) displays the ATLASGAL 870~$\mu$m emission at contour levels starting from $5\times, 10\times, 20\times, 40\times,$ and $80\times 1 \sigma$ where $1\sigma = 0.5~$Jy/beam overlaid atop the 3.3~GHz continuum emission along with the synthesised beam of the VLA observations (filled white circle) as well as the central beam of upGREAT/SOFIA (red) and the LABOCA/APEX beam (teal), for comparison. The position from which the spectra are extracted is marked by the beam in black and labelled C, we also mark the positions of G34.26+0.15~A, B, and D. Panels (b), (c), and (d) display the integrated intensity contours of the 3.264~GHz, 3.335~GHz, and 3.349~GHz CH lines, respectively, overlaid on top of the 3.3~GHz continuum emission. The intensities are integrated over a velocity range between range between 57 and 64~km~s$^{-1}$. The contour levels mark $5\times, 10\times, 20\times, 40\times, 80\times,$ and 160$\times 1 \sigma$ where $1\sigma = 2.65$~K~km~s$^{-1}$ for all three of the CH ground state lines, respectively (solid dark blue curves).} 
    \label{fig:G34P26_map}
\end{figure*}

\begin{figure*}
    \centering
    \includegraphics[width=1\textwidth]{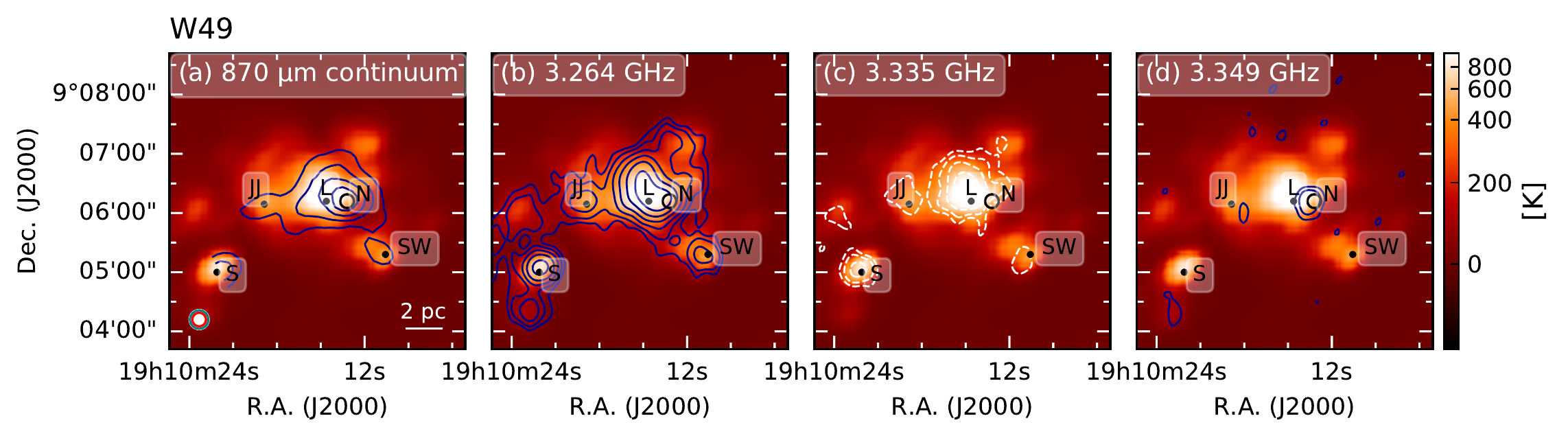}
    \caption{Overview of the 3.3~GHz CH line observations toward W49. Panel (a) displays the ATLASGAL 870~$\mu$m emission at contour levels starting from $5\times, 10\times, 20\times, 40\times,$ and 80$\times 1 \sigma$ where $1\sigma = 1~$Jy/beam overlaid atop the 3.3~GHz continuum emission along with the synthesised beam of the VLA observations (filled white circle) as well as the central beam of upGREAT/SOFIA (red) and the LABOCA/APEX beam (teal), for comparison. The position from which the spectra are extracted is marked by the beam in black and labelled N, we also mark the positions of W49~L, JJ, S, and SW as per \citet{depree1997}. Panels (b), (c), and (d) display the integrated intensity contours of the 3.264~GHz, 3.335~GHz, and 3.349~GHz CH lines, respectively, overlaid on top of the 3.3~GHz continuum emission. The intensities are integrated over a velocity range between range between 2 and 20~km~s$^{-1}$ except in the case of the 3.349~GHz line for which the emission component was integrated between 7 and 14~km~s$^{-1}$. The contour levels mark $5\times, 10\times, 20\times, 40\times, 80\times,$ and $160\times 1 \sigma$ where $1\sigma = 6.08,$ 1.14, and 1.14~K~km~s$^{-1}$ for the 3.264~GHz, 3.335~GHz, and 3.349~GHz CH emission, respectively (solid dark blue curves) and the $3\times, 4\times,$ and 5$\times 1\sigma$ levels where $1 \sigma = 2.28$~K~km~s$^{-1}$ for the 3.335~GHz absorption component (dashed white curves).
    }
    \label{fig:W49_map}
\end{figure*}

\begin{figure*}
    \centering
    \includegraphics[width=1\textwidth]{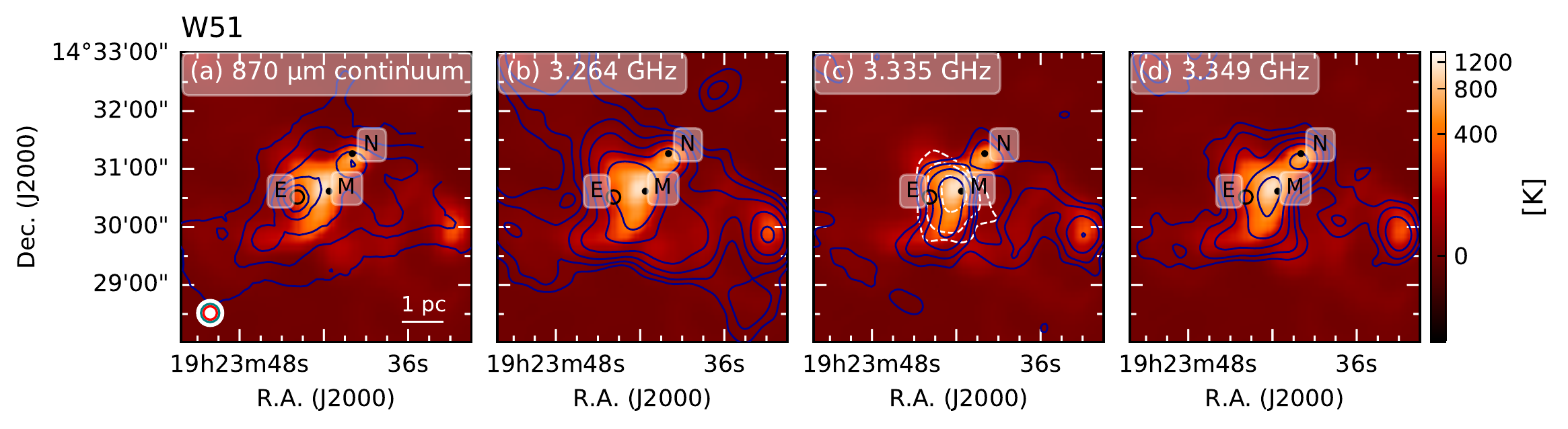}
    \caption{Overview of the 3.3~GHz CH line observations toward W51. Panel (a) displays the ATLASGAL 870~$\mu$m emission at contour levels starting from $5\times, 10\times, 20\times, 40\times,$ and $80\times 1 \sigma$ where $1\sigma = 0.5~$Jy/beam overlaid atop the 3.3~GHz continuum emission along with the synthesised beam of the VLA observations (filled white circle) as well as the central beam of upGREAT/SOFIA (red) and the LABOCA/APEX beam (teal), for comparison. The position from which the spectra are extracted is marked by the beam in black and labelled E, we also mark the positions of W49~M and N. Panels (b), (c), and (d) display the integrated intensity contours of the 3.264~GHz, 3.335~GHz, and 3.349~GHz CH lines, respectively, overlaid on top of the 3.3~GHz continuum emission. The intensities are integrated over a velocity range between range between 61 and 70~km~s$^{-1}$, except in the case of the 3.335~GHz line whose absorption component is integrated between 60 and 66.6~km~s$^{-1}$, while the emission between 64 and 72~km~s$^{-1}$. The contour levels mark $5\times, 10\times, 20\times, 40\times, 80\times,$ and 160$\times 1 \sigma$ where $1\sigma = 2.80, 0.97,$ and 2.80~K~km~s$^{-1}$ for the 3.264~GHz, 3.335~GHz, and 3.349~GHz CH emission, respectively (solid dark blue curves) and the $3\times, 4\times,$ and 5$\times 1\sigma$ levels where $1\sigma = 2.31~$K~km~s$^{-1}$ for the 3.335~GHz absorption component (dashed white curves).}
    \label{fig:W51_map}
\end{figure*}

In Figs.~\ref{fig:SgrB2_G34P26_spec} and \ref{fig:W49N_W51_spec} we present the calibrated and baseline subtracted spectra of all three ground state HFS transitions of CH near 3.3~GHz toward Sgr~B2~(M), G34.26+0.15, W49~(N), and W51~E, respectively. Alongside these ground state lines we also display the corresponding HFS deconvolved spectra of the CH $N,J = 1,1/2 \rightarrow 2,3/2$ transitions near 2006~GHz presented in \citet{wiesemeyer2018unveiling} and \citet{jacob2019fingerprinting} observed using upGREAT/SOFIA. In Figs.~\ref{fig:SgrB2_map} to \ref{fig:W51_map} we compare the distribution of the 3.3~GHz continuum emission and the integrated intensities of the ground state HFS lines of CH (integrated over the velocity dispersion of each source) with that of the emission traced by the cold, dense dust at 870~$\mu$m observed as a part of the ATLASGAL survey \citep{Schuller2009}.

\subsection{Spectral line profiles}\label{subsec:spec_profiles}
The $N,J\!=\!2,3/2\!\rightarrow\!1,1/2$ CH spectrum shows deep absorption profiles not only at velocities corresponding to the LOS features arising from the different spiral-, and inter-arm crossings but also toward the envelope of the molecular clouds. However, the lower satellite line of the ground state is always seen in emission, both toward the molecular cloud as well as along the LOS, while both the main and upper satellite lines show either absorption, emission or a mixture of both. Surprisingly, in the case of the latter two lines, we do not detect significant absorption or emission along the LOS which makes it difficult to constrain the physical conditions of these regions (except toward Sgr~B2~(M)). 

Furthermore, we observe a blend of several components toward the systemic velocities of these sources which can be attributed to different foreground, and background layers of the envelope surrounding these regions with possible contributions also arising from nearby H{\small II} regions, as well. Therefore, prior to carrying out our non-LTE analysis, we first decompose the contributions from the different cloud components associated with each of the molecular clouds. A similar decomposition of the envelope components was carried out for the sight lines toward Sgr~B2 and W51 by \citet{Genzel1979} for the CH ground state lines observed using the Effelsberg 100~m telescope with a 4$^{\prime}$ Gaussian beam. We characterise the different components using Gaussian profiles with velocities and line widths set by using the spectra of the $N,J\!=\!1,3/2\rightarrow\!1,1/2$ HFS transitions of CH near 532~GHz presented in \citet{gerin2010interstellar}, \citet{Qin2010} or retrieved from the Herschel archives\footnote{See, \url{http://archives.esac.esa.int/hsa/whsa/}} pointed toward the same positions as our study, as templates. The spectra of the 532~GHz lines of CH toward our sample of sources are quite complex, showing a combination of foreground absorption and emission corresponding to the velocity of the molecular cloud. These velocities are further cross-checked using position-velocity diagrams and the line widths were compared to the line profile properties of other related interstellar species like CN. Additionally, the spectral line properties are derived by fitting Gaussian components, simultaneously to all three of the ground state lines of CH such that the intrinsic line width and velocities are consistent with one another. The Gaussian fits to the 2006~GHz lines represent the same components from which the column densities are determined, which will be later used as an important constraint in our non-LTE analysis. The subsequently derived peak intensities, FWHM and velocities with respect to $\upsilon_{\rm LSR}$ for the 3.3~GHz lines are tabulated in Table~\ref{tab:derived_line_params}. In the following sections we present a qualitative description of the observed line profiles toward each source. 

\begin{table*}
\centering 
        \caption{Derived line properties. The columns are from left to right: the source designation, transition frequency, LSR velocity, FWHM of the line, peak $T_{\rm B}$, the corresponding line ratio $R_{i}, i=1,2$ where $R_{1} = T_{3.264~{\rm GHz}}/T_{3.349~{\rm GHz}}$ and $R_{2} = T_{3.335~{\rm GHz}}/T_{3.349~{\rm GHz}}$ and the gas density, and temperature derived from the non-LTE radiative transfer modelling.}
    \begin{tabular}{l l rrr c c l c l}
    \hline
    \hline 
         Source & Transition & \multicolumn{1}{c}{$\upsilon$} & \multicolumn{1}{c}{$\Delta \upsilon$} & \multicolumn{1}{c}{$T_{\rm B}$} & \multicolumn{1}{c}{$R_{i}$} &  
         log$_{10}(n/{\rm cm}^{-3})$ & \multicolumn{1}{c}{$T_{\rm kin}$}\\
         & [GHz] & \multicolumn{1}{c}{[km~s$^{-1}$]} & \multicolumn{1}{c}{[km~s$^{-1}$]} & \multicolumn{1}{c}{[K]} &  & & \multicolumn{1}{c}{[K]} 
         \\ 
         \hline 
         Sgr~B2~(M) & 3.264 & $-$58.4(0.1) & 5.4(0.7) & 6.9(1.2) & 1.7 
         &$<2.12$& $<93.0$ \\
                   & 3.335 & $-$58.5(0.2) & 5.2(0.5) & 3.2(1.0)& 0.8
                   \\
                   & 3.349 & $-$58.4(0.1) & 4.6(1.0) & 4.0(0.8) & 
                   \\
                   & 3.264 & 8.2(0.0) & 2.4(0.3) & 6.4(1.0) & 0.8 
                   & $1.97^{+0.17}_{-0.20}$ & 63.8$^{+8.7}_{-8.7}$ \\
                   & 3.335 & 8.3(0.0) & 2.6(0.2) & 9.6(1.4) & 1.3 
                   \\
                   & 3.349 & 8.3(0.0) & 2.2(0.2) & 7.4(0.8) & 
                   \\
                   & 3.264 & 58.2(0.2) & 15.5(2.0) & 1.2(0.9)& 0.5 
                   & 3.78$^{+0.24}_{-0.25}$ & 124.5$^{+4.5}_{-2.4}$\\
                   & 3.335 & 57.6(0.2) & 15.1(1.4) & 3.7(0.6) & 1.4 
                   \\
                   & 3.349 & 57.5(0.3) & 13.5(1.6) & 2.6(0.6) & 
                   \\
                  
                   & 3.264 & 68.6(0.1) & 17.6(1.2) & 24.6(2.0) & 8.0
                   &3.03$^{+0.14}_{-0.14}$ & 56.3$^{+4.2}_{-4.2}$\\
                   & 3.335 & 66.8(0.2) & 17.9(1.2) & -2.8(0.8) & 0.9 
                   \\
                   & 3.349 & 66.7(0.3) & 15.8(2.6) & -3.0(0.5) & 
                   \\
                   & 3.264 & 75.6(0.5) & 14.2(0.5) & 6.8(0.9) & 3.2 
                   & 4.43$^{+0.32}_{-0.33}$& 85.4$^{+4.5}_{-4.6}$\\
                   & 3.335 & 76.4(0.2)& 12.5(1.4) & -3.2(0.7) & 1.5 
                   \\
                   & 3.349 & 75.6(0.4) & 12.5(2.1) & -2.1(0.5) & 
                   \\  
         G34.26+0.15 & 3.264 & 59.6(0.1) & 3.7(0.7) & 13.4(1.5) & 2.7
         & 4.44$^{+0.40}_{-0.42}$ & 95.5$^{+9.2}_{-9.3}$  \\
                     & 3.335 & 58.5(0.5) & 3.0(0.3) & -1.3(0.8) & 0.3 
                     \\
                     & 3.349 & 59.4(0.1) & 3.1(0.3) & 5.0(0.8) & 
                     \\
                     & 3.264 & 61.4(0.0) & 3.5(0.4) & 35.8(1.6) & 4.9 
                     & 4.27$^{+0.31}_{-0.32}$ & 85.4$^{+2.9}_{-3.0}$  \\
                     & 3.335 & 61.5(0.0) & 3.8(0.2) & 15.5(1.0)& 2.1 
                     \\
                     & 3.349 & 61.0(0.0) & 4.0(0.4) & 7.3(0.6) & 
                     \\
         W49~(N)   & 3.264 & 12.3(0.1) & 5.4(0.2) & 30.4(0.9) & 7.2 
         & 3.85$^{+0.30}_{-0.32}$ & 52.5$^{+2.8}_{-2.9}$\\
                  & 3.335 & 11.9(0.4) & 6.1(0.9) & -3.0(0.7) & 0.7
                  \\
                  & 3.349 & 11.8(0.1) & 5.6(0.2) &  4.2(0.6) & 
                  \\
                  & 3.264 & 15.0(0.0) & 2.2(0.2) & 34.7(1.1) & 19.3 
                  & 3.13$^{+0.14}_{-0.14}$& 82.2$^{+5.6}_{-5.7}$\\
                  & 3.335 & 15.3(0.0) & 1.9(0.3) & 5.4(0.8) &  3.0 
                  \\
                  & 3.349 & 15.1(0.7) & 1.7(0.6) & -1.8(0.7) & 
                  \\
         W51~E    & 3.264 & 65.2(0.1) & 5.7(0.3) & 28.0(1.0) & 4.0 
         & $\sim\!3.48$& 104.3$^{+3.2}_{-3.3}$\\
                  & 3.335 & 64.5(0.5) & 6.1(0.5) & -7.1(1.4) & 1.0 
                  \\
                  & 3.349 & 65.3(0.1) & 5.4(0.5) & 7.3(1.2) & 
                  \\
                  & 3.264 & 66.7(0.6) & 7.7(0.8) & 7.3(1.7) & 2.1 
                  & $<2.64$& $<98.0$\\
                  & 3.335 & 67.0(0.1) & 8.0(1.6) & 6.7(2.3) & 1.9 
                  \\
                  & 3.349 & 66.7(0.1) & 7.1(0.5) & 3.3(1.5)&
                  \\
         
         \hline 
    \end{tabular}

    \label{tab:derived_line_params}
\end{table*}

\subsection*{Sgr~B2~(M)}
Located close to the Galactic centre (GC), at a heliocentric distance of ${\sim\!8.15}~$kpc \citep{Reid2019}, the LOS toward the Sgr~B2 giant molecular cloud (GMC) intersects several spiral-arm and inter-arm clouds. The most prominent features arise from the Sagittarius, Scutum, 4~kpc, and 3~kpc arms at LSR velocities of $+8$, $+22$, $-22$, and $-45~$km~s$^{-1}$, respectively, with contributions from clouds near the GC (at a distance \textless2~kpc from it) at velocities near 0~km~s$^{-1}$ \citep{Greaves1994}. LSR velocities between $\approx +60$ and $+70$ \kms\ indicate material associated with the Sgr~B2 region itself. 

The Sgr~B2 GMC contains two main protoclusters hosting a prodigious degree of star formation activity --and has been termed a `mini-starburst' -- namely, Sgr~B2 North (N) and Sgr~B2 Main (M) which are themselves surrounded by a number of smaller hot cores \citep{Bonfand2017, Sanchez2017} and H{\small II} regions \citep[][and references therein]{Depree2015}, all embedded in a ${\sim\!19}~$pc wide low-density envelope (${\sim\!10^{3}~\text{cm}^{-3}}$; \citealt{Schmiedeke2016}). Marked in the 3.3~GHz continuum emission maps presented in Fig.~\ref{fig:SgrB2_map} are the positions of the Sgr~B2~(M) and Sgr~B2~(N) hot cores along with that of a position North-East of (M) labelled (L) which corresponds to an ultra compact (UC) H{\small II}. The FIR transitions of CH near 2006~GHz were observed toward the Sgr~B2~(M) position, which hosts, compared to (N), stronger continuum emission at sub-mm wavelengths, ideal for absorption spectroscopy. Therefore, the analysis of the ground state HFS lines of CH, carried out in the remainder of this paper concentrates exclusively on the Sgr~B2~(M) position. From panel (a) in Fig.~\ref{fig:SgrB2_map} we see that both the 3.3~GHz continuum as well as the CH line emission (and absorption) distributions are offset from the 870~$\mu$m dust emission peaks which traces the densest parts of the H{\small II} region, which corroborates the association of CH with cloud populations of lower gas densities. 

The spectra of all three of the CH ground state transitions show features arising from the environment of the Sgr~B2~(M) cloud near ${\upsilon_{\rm lsr}\!=\!+62}$~km~s$^{-1}$ and also, most distinctively, at velocities corresponding to the GC clouds and the Sagittarius spiral arm. Most remarkably the lower satellite line of CH at 3.264~GHz shows emission features across the entire LOS almost mimicking the wide band absorption seen in the HFS deconvolved spectra of CH at 2006~GHz (see left-hand panel of Fig.~\ref{fig:SgrB2_G34P26_spec}). This, along with the absence of clear CH features ($\geq\!3\sigma$ of the noise level), at LOS velocity components (other than those discussed above), for both the main and upper satellite lines suggests that the lower satellite line is excited such that it shows enhanced emission uniformly across the different velocities covered by this sight line. 

Upon comparing the spectral line profiles toward all three ground state HFS lines toward the envelope of this region, we see that there are three main components at $\upsilon_{\rm lsr} \! = \! +58, +67,$ and $+75$~km~s$^{-1}$ likely tracing extended molecular gas in front of the dense molecular core. Previous investigations by \citet{Sato2000} and \citet{Neill2014} toward the Sgr~B2~(M) region studied using different chemical species have already shown the envelope to comprise of different components at similar velocities. While the +67~km~s$^{-1}$ feature dominates the emission and absorption seen toward this source in the spectra of the lower ground state satellite line and the 2006~GHz FIR line, respectively, it has almost no contributions, with weak absorption if any toward the main and upper satellite lines of the CH ground state. The envelope component associated with $\upsilon_{\rm lsr}\!\sim\!+58~$km~s$^{-1}$ is seen in emission for all three of the HFS components, with the emission toward the lower satellite line being the weakest. In contrast, the $+75~$km~s$^{-1}$ feature shows absorption in the main and upper satellite line.

\subsection*{G34.26+0.15}
At a distance of 1.6~kpc assuming that it belongs to the same cloud complex as G34.43+0.24 \citep{Kurayama2011} the G34.26+0.15 SFR, has the shortest LOS in comparison to any of the sources in our current study. Lying almost tangential to the Sagittarius spiral arm, the main LOS features along this sight line correspond to those of the near-, and far-side crossings of the Sagittarius arm at +11 and +26~km~s$^{-1}$, respectively. At radio continuum frequencies, this region displays two UCH{\small II} regions named A and B, as well as a position C which presents the prototypical example of a H{\small II} region with a cometary morphology \citep[][and references therein]{Reid1985, Heaton1989}. Additionally there also exists an extended 1$^{\prime}$ diameter ring-like H{\small II} region referred to as component D \citep{Reid1985}, which at ${\sim\!3.3~}$GHz shows the dominant continuum emission (see Fig.~\ref{fig:G34P26_map}). The CH spectra are extracted from the position corresponding to component C of G34.26+0.15 but since our observations do not resolve the UCH{\small II} regions A and B from C, the beam and the subsequently extracted spectrum both contain contributions from these regions as well. 

The systemic velocities associated with this region lie between $\upsilon_{\rm lsr} {\sim +57}$ to +66~km~s$^{-1}$. We recognise two main components at these velocities, one centred at the intrinsic velocity of the source near +58~km~s$^{-1}$ and one near $+61.3~$km~s$^{-1}$ which through studies of NH$_{3}$ has been shown to trace infalling cloud layers \citep{Wyrowski2012}. The infalling cloud component is seen in emission in the case of all three HFS transitions while the component closer to the systemic velocity shows weak absorption in the main ground state line. The spectrum of the lower satellite line further shows emission features at +11, and +51~km~s$^{-1}$ respectively, the nature (emission/absorption) of whose corresponding features in the main, and upper satellite lines is difficult to discern down to a 3$\sigma$ noise level.

\subsection*{W49~(N)}
One of the most luminous, ${L\!\sim\!10^{7.2}L_{\odot}}$, \citep{Sievers1991} SFRs in our Galaxy, W49~A, harbours several massive stellar clusters. It contains three massive SFRs, the most prominent of which is W49 North (N). Also termed a mini-starburst, the W49~(N) region hosts several UC~H{\small II} regions arranged in a partial ring-like structure ($\sim$2~pc diameter) \citep{Dreher1984, Welch1987}. We label the different radio peaks following the nomenclature given by \citet{depree1997, depree2020}, who identified their positions using continuum emission maps at 7~mm, 1.3~cm and 3.6~cm. In addition to the position of W49~(N) we also mark the positions labelled L, JJ, W49~South (S) and South-West (SW) each of which represent the H{\small II} regions covered in the 3.3~GHz continuum emission maps presented in Fig.~\ref{fig:W49_map}, where the continuum emission peaks at position L which lies closest to the nominal position of the W49~A GMC. 

Located at a distance of 11.4~kpc \citep{zhang2013parallaxes}, the sight line toward W49~(N) intersects a large amount of foreground material. In addition to the local gas arising from the Perseus arm between $-10$ and +30~km~s$^{-1}$, there are two prominent features at $\upsilon_{\rm lsr}\!=\!+40$~km~s$^{-1}$, and +60~km~s$^{-1}$ corresponding to the near- and far-side crossings of the Sagittarius spiral arm, respectively, as seen in the HFS deconvolved spectrum of CH at 2006~GHz displayed in the left-hand panel of Fig.~\ref{fig:W49N_W51_spec}. The ground state lines show two velocity components at velocities associated with the envelope of this region, redshifted from the systemic velocity of the source at +8.6~km~s$^{-1}$, near +12 and +15~km~s$^{-1}$, respectively. These narrow components arise from the fragmented clumps associated with dense molecular gas of the W49~(N) region as well as from neighbouring infrared star clusters as discussed in \citet{Serabyn1993} and \citet{Alves2003}. The +12~km~s$^{-1}$ velocity component shows absorption only in the main ground state line of CH, while the nature of the +15~km~s$^{-1}$ component in the upper satellite line, seemingly showing weak absorption, is difficult to discern. The lower satellite line of the CH ground state, similar to the sight lines toward Sgr~B2~(M) and G34.26+0.15, shows LOS emission features toward W49~(N) as well.

\subsection*{W51}
The W51~A GMC cloud complex, lying in the Sagittarius spiral arm at a distance of 5.4~kpc \citep{sato2010} has in addition to its massive protocluster, W51 Main (M) (IRS~1), two luminous condensations harboring high-mass young stellar objects present in this region, W51~North (N), also known as W51 IRS~2, and W51 East (E) which constitutes a number of compact hot molecular cores, termed e1--e8, that host hyper- or UC-H{\small II} regions \citep[][and references therein]{sato2010, Ginsburg2017}. Most of the continuum emission in our 3.3~GHz VLA images is dominated by the extended emission of the W51~M region, however we focus on the widely studied W51~E region, toward a nominal position along its ridge (which in itself is not well resolved in our observations), seen as a strong dust continuum peak in Fig.~\ref{fig:W51_map}.   
In the spectrum of the ground state lower satellite line of CH, we see clear emission near +55~km~s$^{-1}$ that can be attributed to the compact H{\small II} region G49.4$-$0.3, with potentially an absorption dip at the same velocities in the case of the main ground state line and no distinct signatures in the upper satellite line. Similar to the case for the CH observations toward W49~(N), we do not see contributions of CH at velocities attributed to the molecular cloud region which for W51~E lies in a velocity interval between $\upsilon_{\rm lsr} = +58~$km~s$^{-1}$ and +63~km~s$^{-1}$. This is not surprising as we expect the CH abundance to peak in more diffuse cloud layers outside of the dense molecular cloud cores, sometimes offset.  The most prominent features we see in our CH spectra of the ground state lines are at +65 and +67~km~s$^{-1}$, respectively. While the former corresponds to the velocity component that is known to typically trace the extended emission along the eastern ridge of the W51~A complex where W51~E resides \citep[see,][]{Carpenter1998}, the latter is associated with the +68~km~s$^{-1}$ molecular clouds that constitute a high velocity stream, first identified in early H{\small I} surveys \citep{Burton1970}. However,  it is unclear whether these two cloud components are physically related or if they represent distinct structures. For the purpose of our studies we treat them as individual components. 

\subsection{CH Column density from the 2006~GHz FIR transition}\label{subsec:coldens_FIR}
The column densities per velocity interval $\text{d}\upsilon$ of the 2006~GHz CH FIR lines can be computed from the deconvolved line profiles expressed in terms of optical depth ($\tau$) versus $\upsilon_{\rm LSR}$ as follows,

\begin{equation}
       \frac{\text{d}N_{u}}{\text{d}\upsilon} = \frac{8\pi\nu^{3}}{c^3A_{\rm E}}\left[ \text{exp}\left(\frac{h\nu}{k_{\rm B}T_{\rm ex}}\right) - 1\right]^{-1}\tau_{\rm decon}(\upsilon) \,  ,
\end{equation}
 with the appropriate spectroscopic parameters $g_{\rm u}$ (the upper level degeneracy), $E_{\rm u}$ (the upper level energy), and $A_{\rm E}$ (the Einstein A coefficient) for every given HFS transition.
The upper level column density in a given velocity range is obtained by integration over $\upsilon$.
The total CH column density $N(\text{CH})$ can be calculated as 
\begin{equation}
  N(\text{CH}) = \frac{N_{u}}{g_{u}} Q~\text{exp}\left( \frac{E_{\rm u}}{T_{\rm ex}}\right),
\end{equation}

\noindent
where the partition function, $Q$, is a function of the rotation temperature, which in LTE is equal to the excitation temperature, $T_{\rm ex}$ (and to the kinetic temperature). 
Typically, the high critical densities of the rotational transitions of many hydrides cause negligible populations of even the lowest rotationally excited levels in the rather low density regions in which these molecules are abundant. This allows their study only with absorption spectroscopy. 
For the case of the CH FIR lines discussed here, we compute critical densities of the order of ${\sim 2\times10^{9}}~$cm$^{-3}$ using collisional rate coefficients calculated by \citet{Dagdigian2018} for gas temperatures between 50 and 100~K. Since, the highest gas densities we would expect along of sight lines are at most 10$^{5}~$cm$^{-3}$ toward the envelopes \citep[see][for Sgr~B2~(M)]{Schmiedeke2016} of the SFRs studied in this work, the CH FIR transitions are sub-thermally excited whereby $T_{\rm ex}$ is lower than the gas kinetic temperatures.  Computed for a two-level system, the value of the critical density is regarded only as an indication of the densities above which collisional processes dominate over radiative ones. In the low density limit ($\leq$ few $\times10^{2}$~cm$^{-3}$) where the level populations are in equilibrium with the cosmic microwave background (CMB) at a temperature $T_{\rm CMB} = 2.73~$K, it is valid to assume that $T_{\rm ex} = T_{\rm CMB}$ or 3.1~K taking into account an estimated contribution from the interstellar radiation field (ISRF) \citep{gerin2010interstellar}. In Table~\ref{tab:column_dens} we summarise the values of the derived column densities computed over the velocity intervals most relevant for the following analysis. Our estimates for the values of $N$(CH) are in agreement with those previously determined by \citet{wiesemeyer2018unveiling, jacob2019fingerprinting}. Note the caveat regarding the determination of the continuum level of the 2006~GHz line spectrum of CH toward W49~(N) as discussed in \citet{jacob2020}.

\begin{table}[]
\caption{Synopsis of the derived CH column densities.}
    \centering
    \begin{tabular}{lrc}
    \hline
    \hline 
         \multicolumn{1}{c}{Source} & $\upsilon_{\rm min}$--$\upsilon_{\rm max}$ & $N$(CH) \\
          & ~[km~s$^{-1}$] & [10$^{14}$ cm$^{-2}$] \\ \hline 
         Sgr~B2~(M)  &-62-- -52 & 0.23 \\
                     & 4--11    & 0.14 \\
                     & 42--90   & 7.28 \\
         G34.26+0.15 & 35--70   & 0.93 \\
         W49~(N)      &  0--25   & 2.21 \\
         W51~E       & 45--72   & 6.62 \\
         \hline 
    \end{tabular}
    
    \label{tab:column_dens}
\end{table}

\subsection{Non-LTE radiative transfer analysis for the CH ground state} \label{subsec:nlte_models}
In an effort to characterise the physical and excitation conditions traced by the CH ground state $\Lambda$-doublet HFS transitions, we perform non-LTE radiative transfer calculations using the radiative transfer code MOLPOP-CEP \citep{Asensio2018}. Based on the (coupled) escape probability formalism presented in \citet{Elitzur2006}, this code provides solutions to the radiative transfer equation for multi-level systems, while also taking into account the effects of line overlap within a plane-parallel slab geometry. Since the code assumes the physical conditions to be uniform across the slab, namely, the gas density, $n_{\rm H}$, the gas kinetic temperature, $T_{\rm kin}$, and the molecular abundance, $X$, it is run in several iterations over varying physical conditions in order to sample different cloud layers. The models were computed over a density-temperature grid of size 100$\times$100, for $n_{\rm H}$ values in the range of 25 to 10$^{5}$~cm$^{-3}$ and gas temperatures between 50 and 175~K. Contributions from the major collision partners, that is, atomic and molecular hydrogen, are weighted according to the molecular fraction, $f_{{\rm H}_{2}}$, of each modelled velocity component. We derive the column-averaged molecular hydrogen fraction using ${f^{N}_{\text{H}_{2}} = 2N(\text{H}_{2})/\left( N(\text{H}) + 2N(\text{H}_{2}) \right)}$ where the values for $N({\rm H})$ are taken from \citet{winkel2017hydrogen} while $N(\text{CH})$ values derived as discussed in Sect.~\ref{subsec:coldens_FIR}, are used as proxies for $N(\text{H}_{2})$, following the relationship between the two molecules as estimated by \citet{sheffer2008}, [CH]/[H$_{2}$] = $3.5^{+2.1}_{-1.4}\times10^{-8}$. We further assume a constant ortho-to-para (OPR) ratio of 3:1 between the ortho and para spin states of H$_{2}$, as expected for thermal equilibrium in the low-temperature limit. The radiative transfer analysis was carried out using HFS resolved rate coefficients of CH excited by collisions with atomic and molecular hydrogen computed by \citet{Dagdigian2018}. These collisional rate coefficients are computed based on high-accuracy calculations of the potential energy surfaces (PESs) of CH--H and CH--H$_{2}$ collisions, for all transitions between the lowest 32 HFS levels for CH($X^2\Pi$) or upper level energies $\leq\!300~$K. They account for both direct as well as indirect excitations with H$_{2}$. More recently \citet{Marinakis2019} have calculated HFS resolved collisional rate coefficients using the most recent ab initio PESs for collisions of CH with He. It is common practice to scale the rate coefficients for collisions with He by a factor equivalent to the square root of the involved reduced mass ratio to obtain the collisional rates with para-H$_2$; the validity of this approximation is limited not only temperature-wise which in turn affects the ortho-to-para ratio but is also dependent on the properties of the PESs used. 
Here we use collisional rate coefficients for collisions with para-H$_{2}$ that are scaled from those of collisions between CH and He along with the direct collisional rate coefficients computed for collisions of CH with atomic hydrogen and ortho-H$_{2}$. We use this combination of collisional rate coefficients because when using the rates that were computed directly for the collisions with para-H$_{2}$ by \citet{Dagdigian2018}, we find that the models are capable of producing level inversion in only the lower satellite line (see Fig.~\ref{fig:other_models}). This might be a direct result of the fact that the CH rate coefficients involving collisions with para-H$_2$ are comparable to those for collisions with ortho-H$_2$. As noted by this author, this is in contrast to the trend observed for collisions of other molecules with H$_2$ for which the collisional cross-sections with ortho-H$_{2}$ are larger than that for para-H$_{2}$, like in related molecules, for example for OH \citep{Schewe2015}, and C$_2$H \citep{Dagdigian2018c2h}. \citet{Dagdigian2018} also find that the fraction of collision-induced transitions that occur through indirect collisions is higher for collisions with para-H$_{2}$ unlike that for ortho-H$_{2}$ for which direct collisions dominate. For the specific case of the ground state transitions this holds true for all HFS transitions except for the $F^{\prime}-F^{\prime\prime}\!=\!1-0$ transition, that is, the upper satellite line, for which the direct contributions dominate at gas temperatures above 100~K. While this parity propensity aids pumping routes based on observations, one would except this to be a prominent effect in the excitation of the lower satellite line.
Therefore, we use the collisional rate coefficients for collisions of CH with para-H$_{2}$ that are scaled from the collisional rate coefficients for collisions of CH with He.

Collisions with charged particles, namely with electrons and heavier ions are not considered in the radiative transfer calculations presented in this paper, but may play an important role in the excitation of the CH $\Lambda$-doublet, particularly in regions with high electron fractions, $x_{e} = n_{e}/(n_{\rm H} + 2n_{{\rm H}_{2}}) >10^{-5}$--$10^{-4}$. Such high electron fractions are the norm in the diffuse molecular clouds present along the lines of sight studied here and may even be prevalent in the clouds surrounding the observed H{\small II} regions themselves, making electrons an important collision partner at low gas temperatures ${T\leq 100}~$K. \citet{Bouloy1977, Bouloy1979} have studied the impact of collisional excitation by electrons on $\Lambda$-doublet transitions with particular emphasis on the ground state $\Lambda$-doublet transitions of OH. They compute the collisional rate coefficients for collisions with electrons, either using perturbation methods such as those used by \citet{Goss1968} or by using the Born approximation, both of which yield comparable results. These authors concluded that collisions with electrons, while incapable of inducing level inversion in the ground state lines of OH at 18~cm, are responsible for thermalising them. \citet{Bouloy1984} have further studied the excitation conditions of the ground state lines of CH. They model the excitation by considering the radiative and collisional (de-)excitation of CH with H, H$_{2}$, and electrons. Their results once again point to the role played by the collisions with electrons in thermalising the CH lines rather than inverting them. However, the excitation temperature of OH is found to be a few 1--2~K above $T_{\rm CMB}$ as derived from the resolved optical spectra of the OH $A-X$ band transitions \citep{Felenbok1996} or when measured by comparing the emission and absorption profiles of the radio L-band transitions of OH \citep[and references therein]{Liszt1996}. This implies that densities much higher than the critical density are needed for thermalisation, this might similarly be the case for CH. More recently, \citet{Goldsmith2017} have examined the impact of electron excitation on high dipole moment molecules like HCN, HCO$^+$, CS and CN in various interstellar environments. Since long-range forces dominate the collisional cross-sections for electron excitation, the cross-sections and, in turn, the collisional rate coefficients scale with the square of the electric dipole moment, $\mu_{\rm e}$. Hence, the electron collisional rate coefficients for CH having $\mu_{\rm e} = 1.46$~D \citep{Phelps1966} can be approximated to be ${\simeq\!25\,\%}$ of those of HCN with $\mu_{\rm e} = 2.985~$D \citep{Ebenstein1984}. Scaling the value of the HCN--e$^{-}$ collisional rate coefficient at $T\leq 100~$K from \citet{Faure2007}, we find the CH--e$^{-}$ collisional rates to be of the order of $\sim\!2\times10^{-5}~$cm$^{3}$~s$^{-1}$. From this we can compute the critical electron fractional abundance, $x^{*}({\rm e}^{-})$, which defines the fractional abundance of electrons required for the collisional rate coefficients with electrons to be the same as that with H$_{2}$ such that $x^{*}({\rm e}^{-}) = n_{\rm crit}(\rm e^{-})/n_{\rm crit}({\rm H}_{2})$. Under the validity of these assumptions, $x^{*}({\rm e}^{-})$ for CH is approximated to ${\sim\!10^{-6}}$, making CH likely to be affected by electron excitation. Therefore, a complete treatment of the radiative--collisional (de-)excitation of the CH ground state would still require the availability of accurate collisional rate coefficients for collisional excitation by electrons. But to our knowledge, these are currently not available. \\

The models for each velocity component uses a fixed line width corresponding to the intrinsic values estimated from the observations of the ground state lines, as detailed in Sect.~\ref{subsec:spec_profiles}. External radiation from the CMB and ISRF \citep{Mathis1983} forms the main sources of radiative excitation present within these models, while the internal radiation is emitted by warm dust grains. The dust temperature used are given by the PACS 160~$\mu$m dust continuum and taken from \citet{Konig2017}. As discussed earlier we find that the effects of radiative excitation alone cannot create a population inversion between the two $\Lambda$-doublet levels of the ground state of CH as the excitation is followed by radiative decay. Therefore, such a simple approximation cannot create the anomalous excitation observed in CH and so we invoke the effects of FIR line overlap. Governed by a fixed set of input parameters, in each iteration the code solves the radiative transfer problem by dividing the slab into several `zones' characterised by increasing optical depths (or column densities). The zones within each slab were constrained by the column density of the velocity component as derived using the ${N, J\!=2,3/2\rightarrow\!1,1/2}$ CH lines, discussed in Sec.~\ref{subsec:coldens_FIR}. The radiative transfer analysis is simplified by considering the distribution of the independent line ratios under the assumption that the contributions from all three of the ground state HFS transitions of CH have the same beam filling factor. In the following sections we discuss the results obtained from the non-LTE radiative transfer analysis.
\begin{figure*}
    \includegraphics[width=12.6cm]{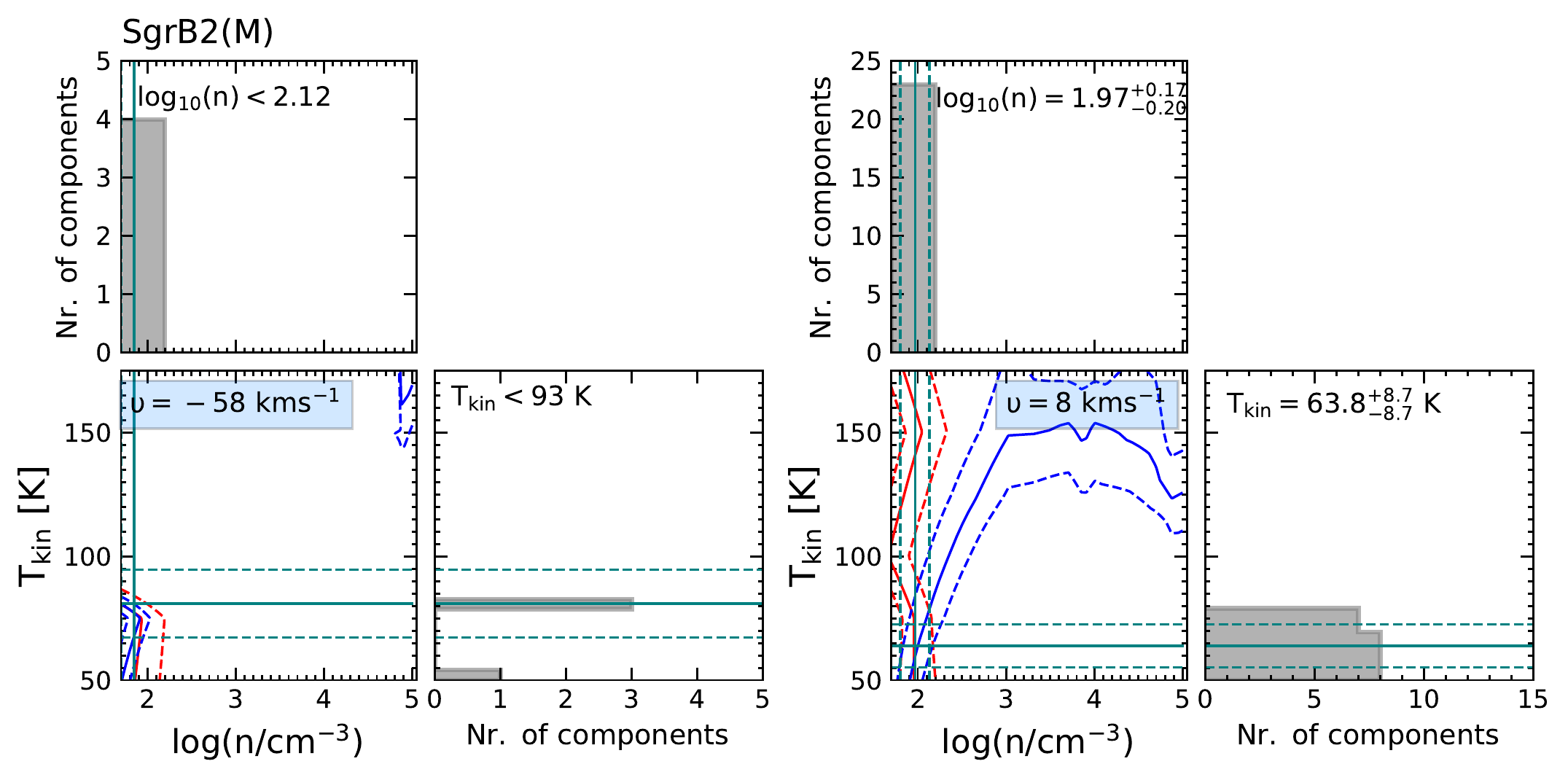} \\
    \includegraphics[width=1.02\textwidth]{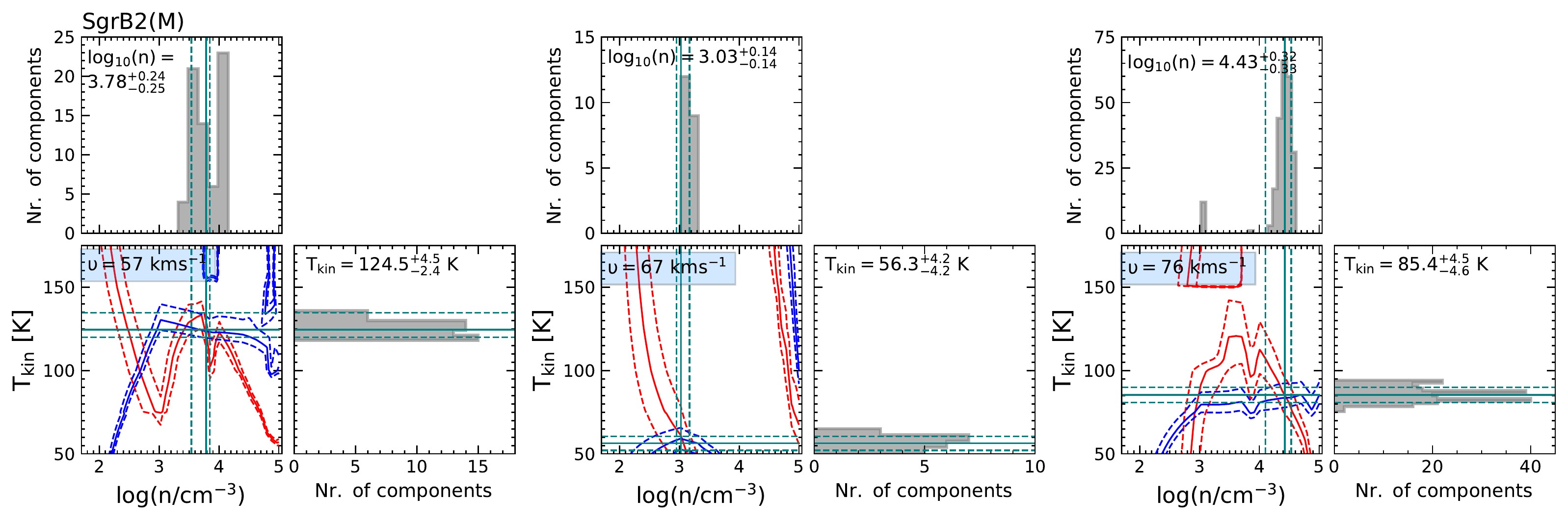}
    \caption{Clockwise from the top left: MOLPOP-CEP non-LTE radiative transfer modelling of the CH ground state transitions toward Sgr~B2~(M) for LOS velocity components at -58, and +8~km~s$^{-1}$ followed by those associated with the envelope at $\upsilon_{\rm LSR}$ = +76, +67, and +57~km~s$^{-1}$, respectively. Displayed in red and blue curves (central plot) are the modelled line ratios which best reproduce the observed line ratios between the 3.264~GHz, and 3.349~GHz lines and those between the 3.335~GHz, and 3.349~GHz lines and their uncertainties, respectively, as a function of $n$ and $T_{\rm kin}$. While the corner plots display the gas density, and temperature distributions as governed by the 3~$\sigma$ level of the minimum $\chi^{2}$ with the median and standard deviation marked by the solid and dashed teal lines, respectively.}
    \label{fig:SgrB2_molpop_results}
\end{figure*}
\begin{figure*}
\sidecaption
    \includegraphics[width=12.6cm]{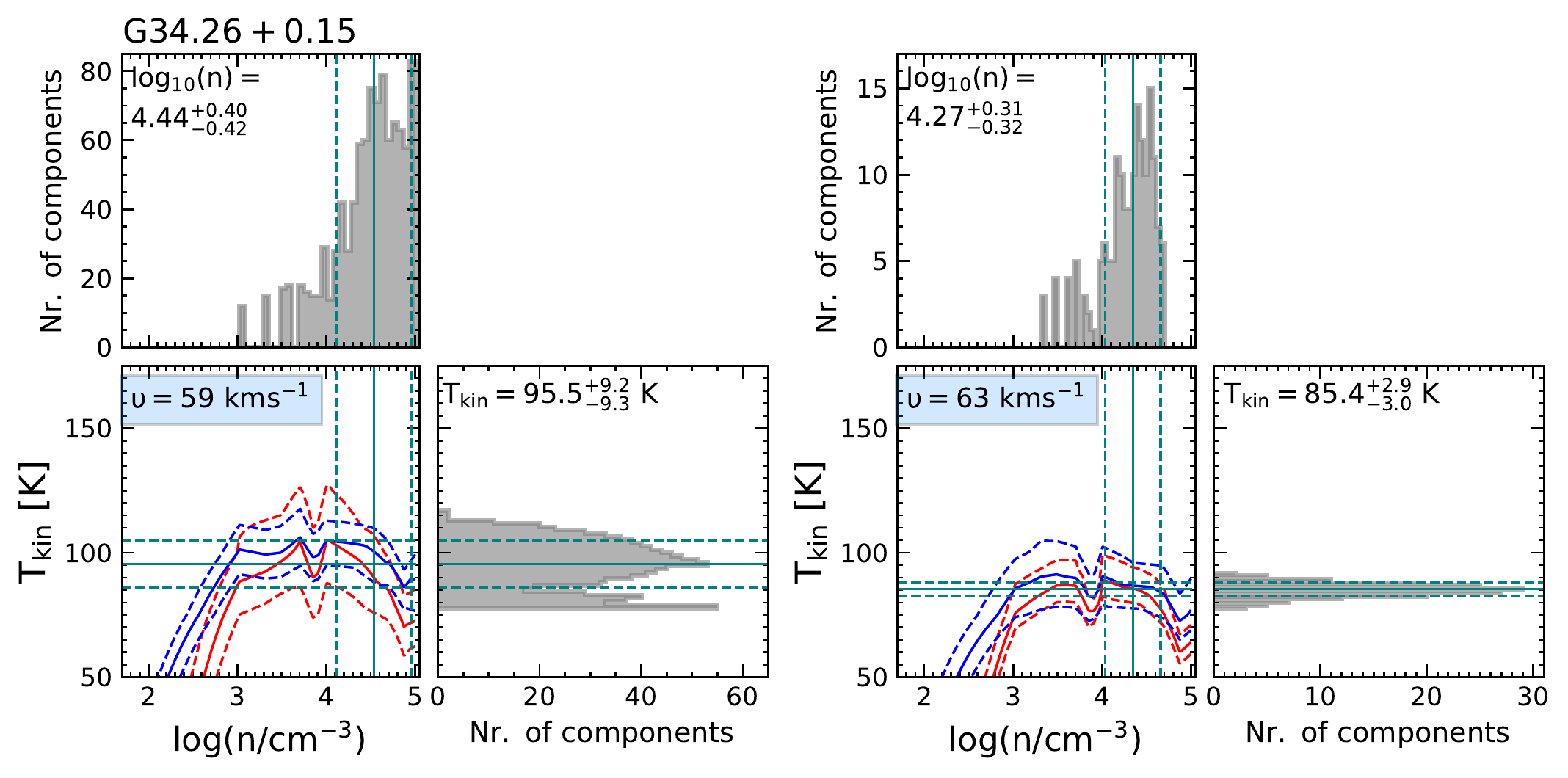}
    \caption{Same as Fig.~\ref{fig:SgrB2_molpop_results} but toward the +59, and +63~km~s$^{-1}$ velocity components of G34.26+0.15.}
    \label{fig:G34P26_molpop_results}
\end{figure*}
\begin{figure*}
\sidecaption
    \includegraphics[width=12.6cm]{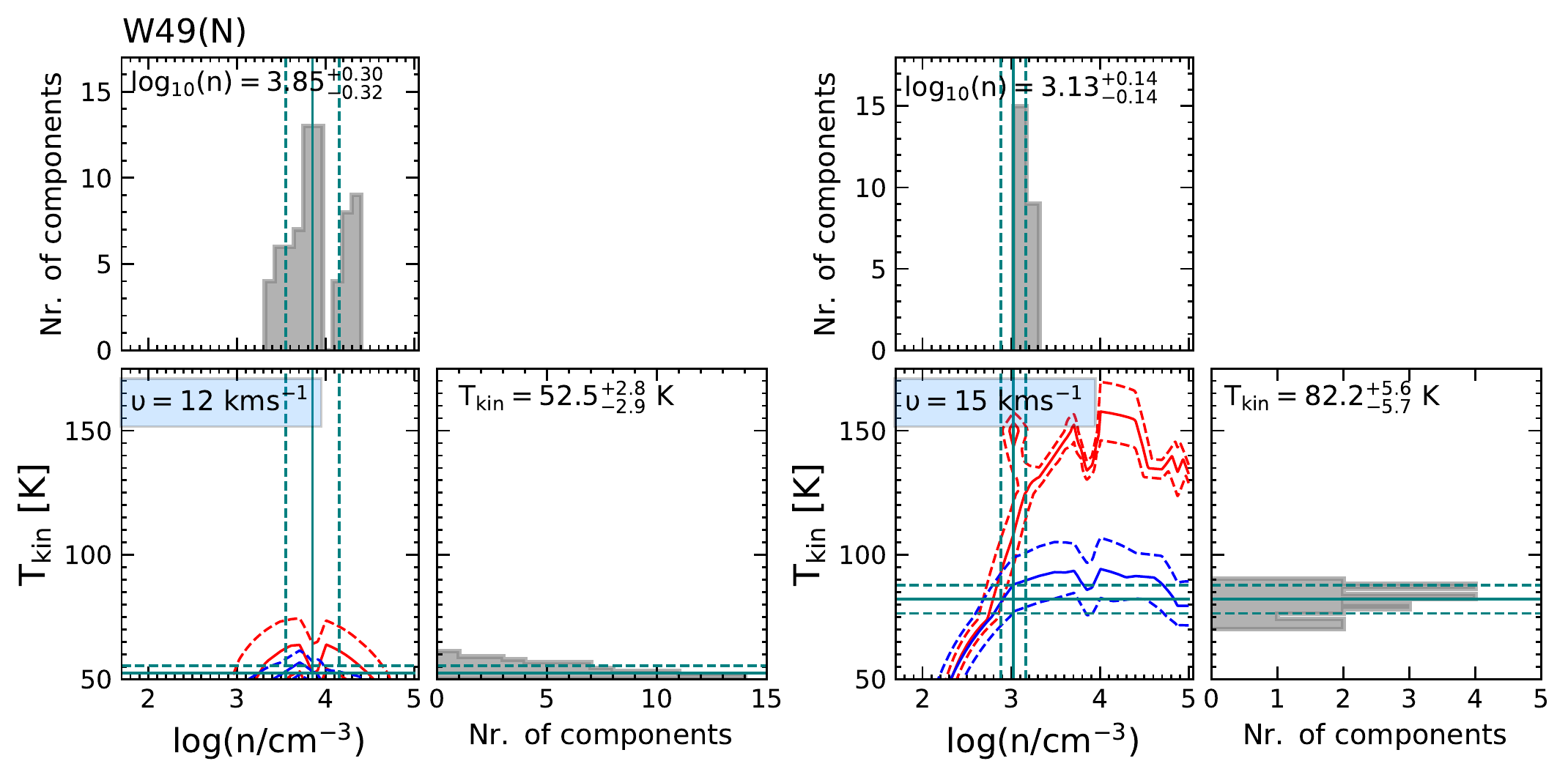}
    \caption{Same as Fig.~\ref{fig:SgrB2_molpop_results} but toward the +12, and +15~km~s$^{-1}$ velocity components of W49~(N).}
    \label{fig:W49N_molpop_results}
\end{figure*}
\begin{figure*}
\sidecaption
    \includegraphics[width=12.6cm]{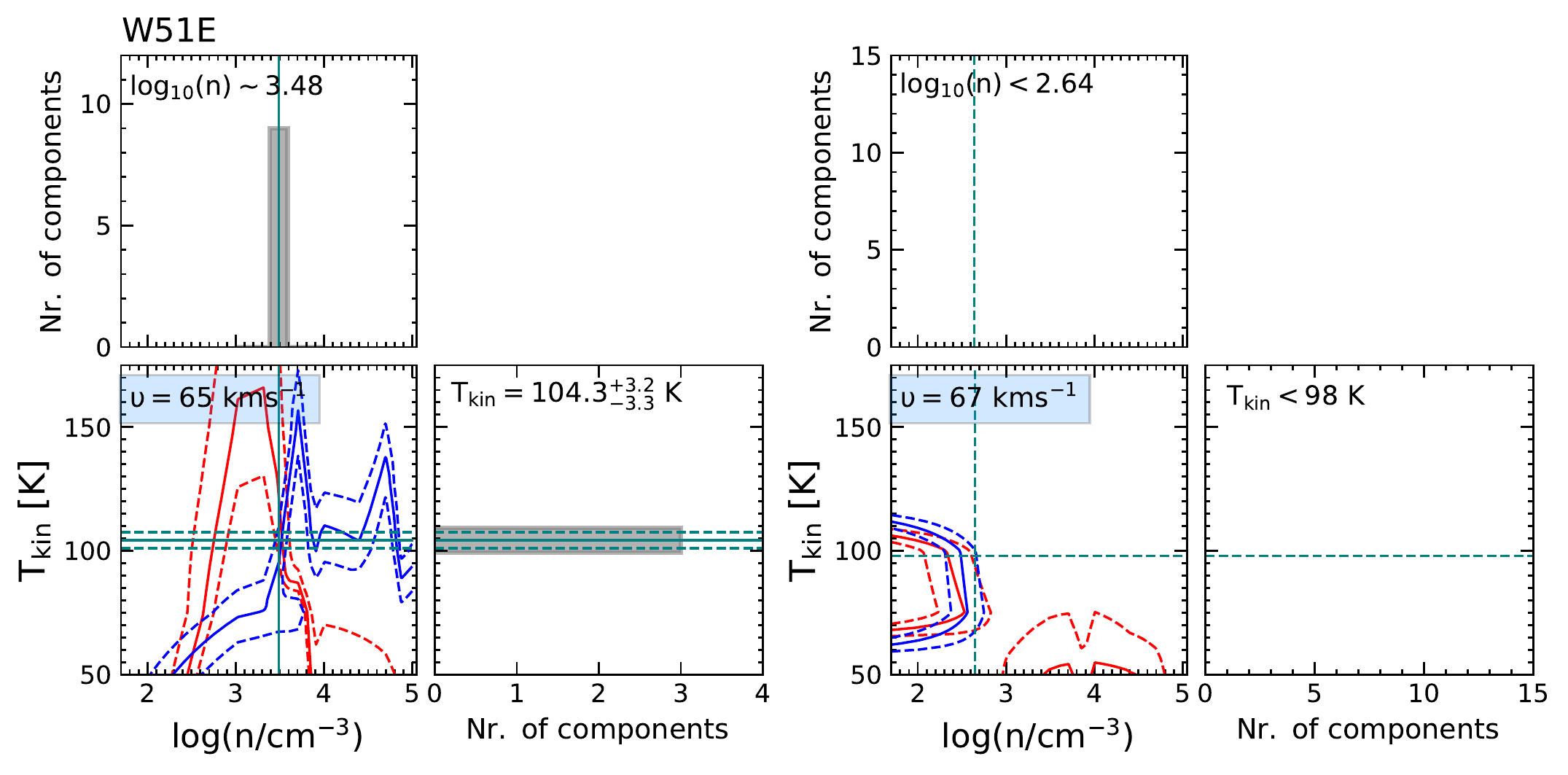}
    \caption{Same as Fig.~\ref{fig:SgrB2_molpop_results} but toward the +65, and +67~km~s$^{-1}$ velocity components of W51~E. Unable to constrain the minimum $\chi^2$ values, the parameter distributions for the velocity component at +67~km~s$^{-1}$ are not displayed here but marked are the upper limits for the gas density and temperature.}
    \label{fig:W51E_molpop_results}
\end{figure*}

\section{Discussion}\label{sec:discussion}
\subsection{Physical conditions of the ground state \texorpdfstring{$\Lambda$}{Lambda}-doublet of CH}\label{subsec:phy_conds}
The physical conditions that prevail in the regions studied by us are constrained based on the behaviour of $\chi^{2}$ across our modelled density-temperature parameter space. The $\chi^{2}$ value is computed across the entire grid as follows,
\begin{align} 
  \chi^{2} &= \Sigma^{2}_{i=1} {\left( R_{\text{i, obs}} - R_{\text{i,mod}} \right)^{2}}/{\sigma^{2}_{R_{\text{i,obs}}}} \, , 
\end{align}
where $R_{\text{i,obs}}$ and $R_{\text{i,mod}}$ represent the ratio of the observed and modelled line brightness temperatures between any two of the three CH HFS lines on the $T_{\text{B}}$ scale, $\sigma^{2}_{R_{\text{i,obs}}}$ represents uncertainties in the line ratios. The independent line ratios used in our analysis are $R_{1} = T_{3.264~{\rm GHz}}/T_{3.349~{\rm GHz}}$ and $R_{2} = T_{3.335~{\rm GHz}}/T_{3.349~{\rm GHz}}$. The modelled solutions are constrained to the 3~$\sigma$ levels of their minimum $\chi^{2}$ distributions. 

We summarise the resulting gas densities, and temperatures for the various velocity components observed toward each source in Table~\ref{tab:derived_line_params}. These results are also visualised across the density-temperature plane in Figs.~\ref{fig:SgrB2_molpop_results} to \ref{fig:W51E_molpop_results}. The models describing the velocity components representing the envelopes of the different SFRs predict optically thin lines under the assumption of physical conditions consistent with what is expected for translucent clouds \citep[see,][]{Snow2006}. Thus our results are consistent with those of previous studies which determine the gas densities and temperatures for these warm extended envelope clouds to be $<10^{5}~{\rm cm}^{-3}$ and $<65$~K, respectively, for example toward the envelope of Sgr~B2~(M) as determined by \citet{Etxaluze2013} using rotational transitions of CO. By weighting the models using a smaller molecular fraction in comparison to that used for the envelope clouds we were able to reproduce the observed line ratios for the velocity components observed toward the LOS features. We do so, for the two LOS features consistently observed in emission in all three HFS lines toward Sgr~B2~(M) at $\upsilon_{\rm LSR} = -58$, and +8.3~km~s$^{-1}$, respectively. In particular for the narrow LOS features, the different zones within the slab models were just able to simulate the values of the CH column density that were constrained by the 2006~GHz CH transition between 5$\times10^{12}$ and $10^{13}~$cm$^{-2}$. The gas densities derived in this work for the LOS components of Sgr~B2~(M) are similar to those derived by \citet{Thiel2019A} using the absorption line studies of other simple molecules like HCO$^+$ whose abundance has been shown to be well correlated with that of CH in diffuse clouds \citep{gerin2010interstellar}. 

Lying just outside of the bow of a cometary-shaped UCH{\small II} region, it is widely believed that the molecular material in G34.26+0.15 is heated by the H{\small II} region \citep[][and references therein]{Mookerjea2007}. Therefore, from the results of our non-LTE radiative transfer modelling it is clear that the CH emission arises from warm molecular layers of moderate gas densities of 2--3$\times10^{4}~$cm$^{-3}$. Compared to G34.26+0.15, for the envelope of W49~(N) we derive lower gas densities and temperatures. 
However, this is not surprising and consistent with the average physical conditions ($n = 10^{4}~$cm$^{-3}$, $T_{\rm kin} = 130~$K) associated with the photo-dissociation region (PDR) of W49~(N) as derived using the [C {\small II}] 158~$\mu$m line by \citet{Vastel2001}. Similarly in the W51 region we trace a warm extended gas present in the W51~A complex rather than the very compact dense e1--e8 hot molecular cores. 

The large observed abundances of CH in the warm envelope layers of SFRs are conceivable because chemical models predict that the formation of CH starts with the endothermic hydrogen abstraction reaction forming CH$^{+}$ ($\Delta E/k_\text{B}$ = 4620~K; \citet{hierl1997rate}) acting as a bottleneck for the subsequent exothermic reactions building up CH$_{2}^+$ and CH$_{3}^+$, in regions with locally elevated gas temperatures as in the case of turbulent dissipation regions \citep[TDRs;][]{Falgarone2005, Godard2009,Godard2012,godard2014chemical}, which is then followed by the dissociative recombination of CH$_{2}^+$ or CH$_{3}^+$ to form CH. Such a dissipation of turbulence may also be responsible for large scale velocity gradients which in turn can cause line overlap between different HFS splitting levels pertaining to the same sub-mm or FIR rotational transition.

Recently, a similar range of gas densities and temperatures were found to be probed by CH$_{2}$, which is also formed from the dissociative recombination of CH$_{3}^+$ \citep{jacob2021}. Given the high energies above the ground state, 225~K, of the 68-71~GHz CH$_2$ lines observed in just two regions, in the past it had been believed that the CH$_{2}$ emission, unlike that of CH, was associated with the dense hot molecular cores of newly formed stars, despite the two molecules sharing the same chemical history. In contrast the recent observations of \citet{jacob2021}, which have extended the detections of this molecule, have convincingly shown that its emission arises from the warm ($T_{\rm kin} = 163~$K) dilute ($n = 3.4\times10^{3}~$cm$^{-3}$) layers of PDRs. Moreover, the ${N_{K_{\text{a}}K_{\text{c}}}=\,4_{04} - 3_{13}}$ transition of CH$_{2}$ between 68 and 71~GHz observed by these authors also shows a weak masing effect in which all three of the fine-structure lines corresponding to this transition also show weak level inversion. Together, the non-LTE analysis carried out for CH$_{2}$ by \citet{jacob2021} and CH here, support the spatial co-existence of the two molecules as well as their chemistry. For example, in the W51 region, both CH and CH$_{2}$ show stronger emission towards the extended H{\small II} region of W51~M rather than the denser and more massive nearby regions of W51~E and W51~N. Additionally, the association of both CH and CH$_{2}$ with TDRs or warm PDR layers may open up formation and destruction routes between the two molecules that would otherwise be unlikely because of the low temperatures of the cold neutral medium (CNM).

\subsection{CH in the cold dark cloud TMC-1 and its neighbourhood}
Overall the physical conditions derived from our non-LTE analysis suggests that CH in the vicinity of the dense SFRs traces the warm translucent gas layers of their envelopes. However, in order to cross-check the validity of our results we additionally model the CH ground state lines observed toward the extensively studied dense, cold and dark Taurus Molecular Cloud-1 (TMC-1), whose kinetic temperature is well known, ${\approx\!10~}$K. As noted by \citet{Rydbeck1976} there are very few examples toward dark clouds in which the relative intensities of the CH ground state lines show departures from LTE. In TMC-1, these authors detected non-LTE behaviour only toward certain positions along the dense ridge that forms the spine of this region.

Based on their CH and OH observations, \citet{Xu2016} suggest that CH traces a C-type shock front in the boundary region between the dense gas in TMC-1 and translucent and diffuse gas. Here, the relative intensities of the CH (and the OH) radio lines indicate anomalous excitation in certain positions along its outer boundary. The observations by \citet{Goldsmith2010} of rotationally excited H$_2$ lines in this environment require temperatures of at least 200~K and modest H$_2$ column densities of $1$--$5~\times10^{18}~$cm$^{-2}$, which, according to these authors, `points to an enhanced heating rate which may be the result of, for example, dissipation of turbulence'.
We suggest that this mechanism, dissipation of turbulence, which in Sect.~\ref{subsec:phy_conds} we have invoked for CH production, may likely also explain the rare instances of anomalous excitation of the CH ground state lines in such environments without strong sub-mm/FIR continuum radiation (and also co-determine their CH abundances).

Concentrating on the dense ridge of TMC-1 itself, we `calibrate' the results with data from previous observations of the CH ground state HFS lines, carried out by \citet{Suutarinen2011} and \citet{Sakai2012}. \citet{Suutarinen2011}, while observing an abundance gradient in CH across the TMC-1 region, do not find significant deviations in the relative intensities of the three ground state HFS lines of CH from their expected LTE line ratio toward much of TMC-1.

\begin{figure}
    \centering
    \includegraphics[width=0.45\textwidth]{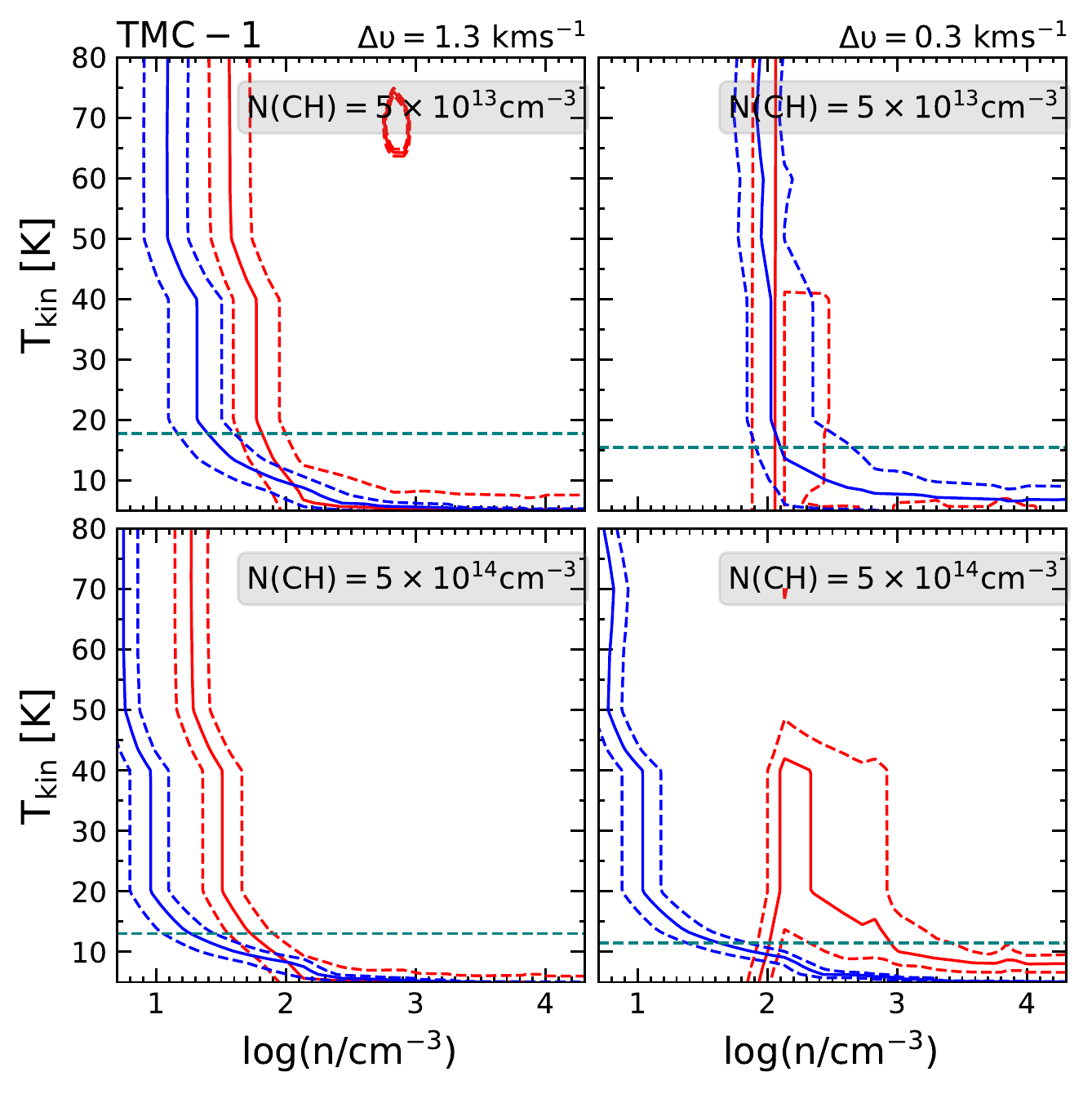}
    \caption{MOLPOP-CEP non-LTE radiative transfer modelling of the CH ground state transitions toward the broad (left) and narrow (right) velocity components observed toward the cyanopolyyne peak in TMC-1 for $N$(CH) = $5\times10^{13}~$cm$^{-2}$ (top panel) and $5\times10^{14}~$cm$^{-2}$ (bottom panel). Displayed in red and blue solid and dashed curves are the modelled line ratios which best reproduce the observed line ratios between the 3.264~GHz, and 3.349~GHz lines and the 3.335~GHz, and 3.349~GHz lines and their uncertainties, respectively, as a function of $n$ and $T_{\rm kin}$. The dashed teal line marks the upper limit for the gas temperatures for the broad and narrow features.}
    \label{fig:TMC1_denstemp}
\end{figure}

Toward the cyanopolyyne peak (CP) in TMC-1, a region that shows enhanced abundances of long carbon chain molecules and other unsaturated C-bearing species \citep{Little1979}, both of the studies discussed above observe two velocity components in the CH spectra, namely, a narrow component ($\Delta\upsilon = 0.3~$km~s$^{-1}$) and a relatively broader component ($\Delta\upsilon = 1.3~$km~s$^{-1}$). We model the observed lines ratios of both velocity components using the same combination of collisional rate coefficients as before but reduce the number of energy levels included in the models by half. We use only the first 16 energy levels here for the case of TMC-1, in order to avoid numerical singularities that may arise when solving the equations of radiative transfer for energy levels as high as 389~K at low gas temperatures of 5~K. We run the models over a $50\times50$ density-temperature grid probing gas densities between 50 and $2.5\times10^{4}$~cm$^{-3}$ and gas temperatures between 5 and 100~K using a slab geometry while also retaining effects of line overlap. We also explore the parameter space for two different values of CH column densities at $5\times10^{13}$ and $5\times10^{14}~$cm$^{-2}$ since the $N$(CH) values quoted by previous studies are determined by assuming $T_{\rm ex}=-60$ to $-20~$K, a range of values that is frequently used, but it in fact holds little justification from observations. The CH ground state line ratios modelled toward TMC-1 across a sub-set of the modelled density-temperature parameter space are displayed in Fig.~\ref{fig:TMC1_denstemp}. Unable to converge to a single local minimum, we are only able to comment on the upper limits of the gas temperature derived toward each of the two velocity components. The upper limits for the gas temperature for the models with $N$(CH)=$5\times10^{13}~$cm$^{-2}$ are 15~K and 18~K, and 11~K and 13~K when $N$(CH)=$5\times10^{14}~$cm$^{-2}$, for the narrow and broad velocity components, respectively. The derived temperatures are consistent with what is expected in the TMC-1 region \citep[see][and references therein]{Sakai2012} while the gas densities display only a lower limit of 130~cm$^{-3}$, which is also in agreement with the interpretation presented by \citet{Sakai2012}, that CH and in particular the broad component traces the less dense envelope material. \\

\subsection{Excitation conditions of the ground state \texorpdfstring{$\Lambda$}{Lambda}-doublet of CH}\label{subsec:exc_conds}
\begin{figure*}
    \includegraphics[width=0.5\textwidth]{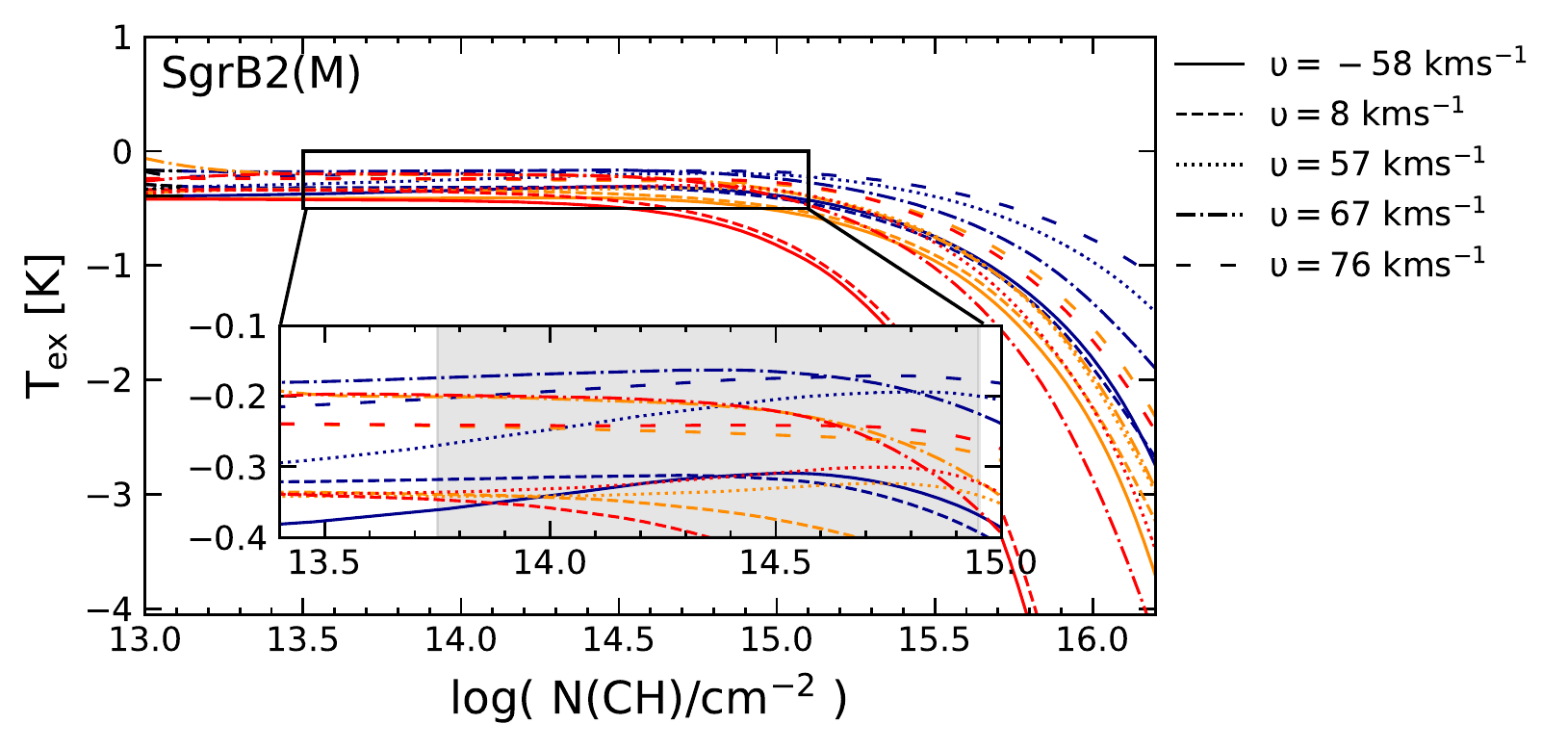}\quad 
    \includegraphics[width=0.5\textwidth]{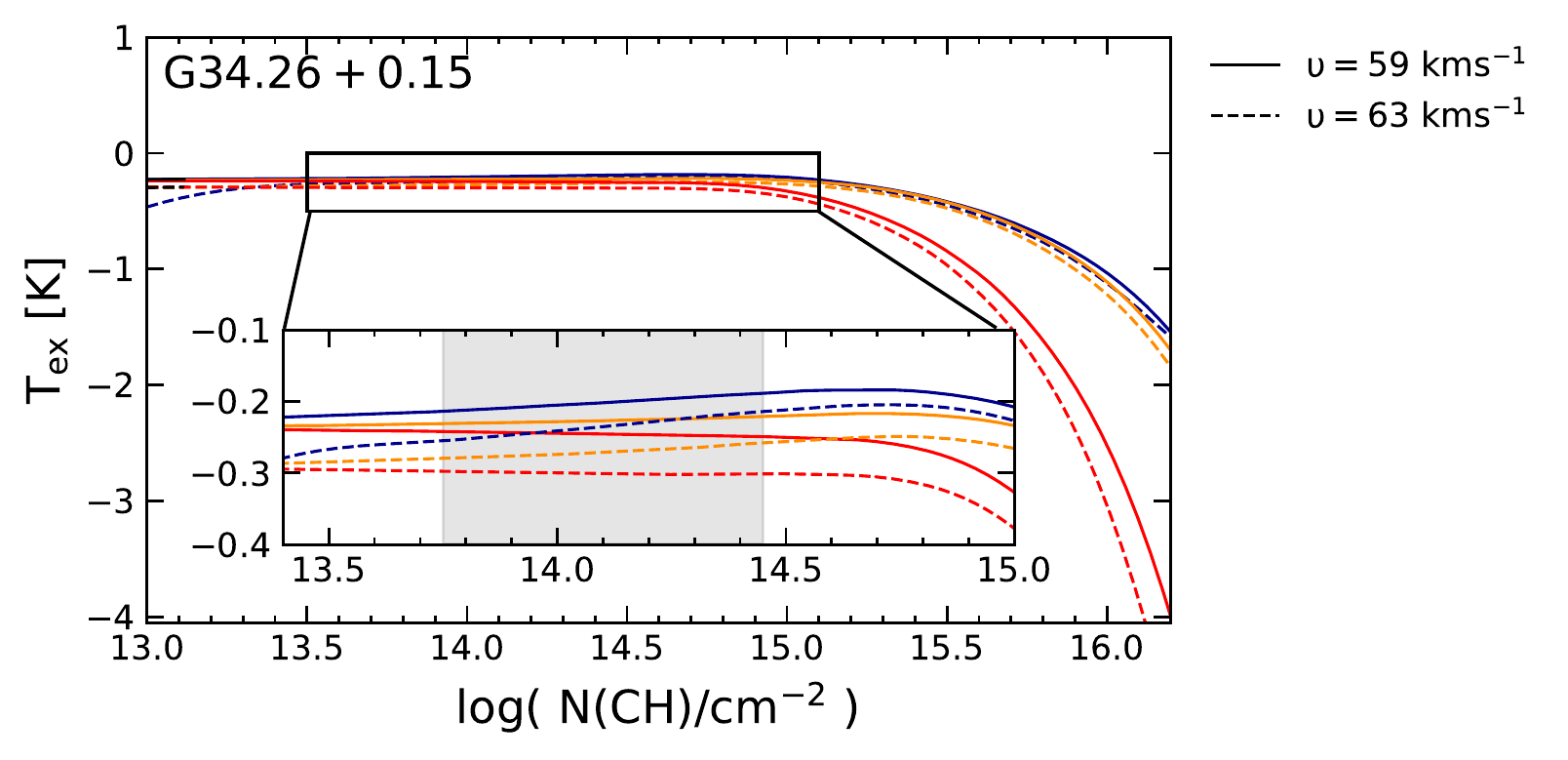}\\
    \includegraphics[width=0.5\textwidth]{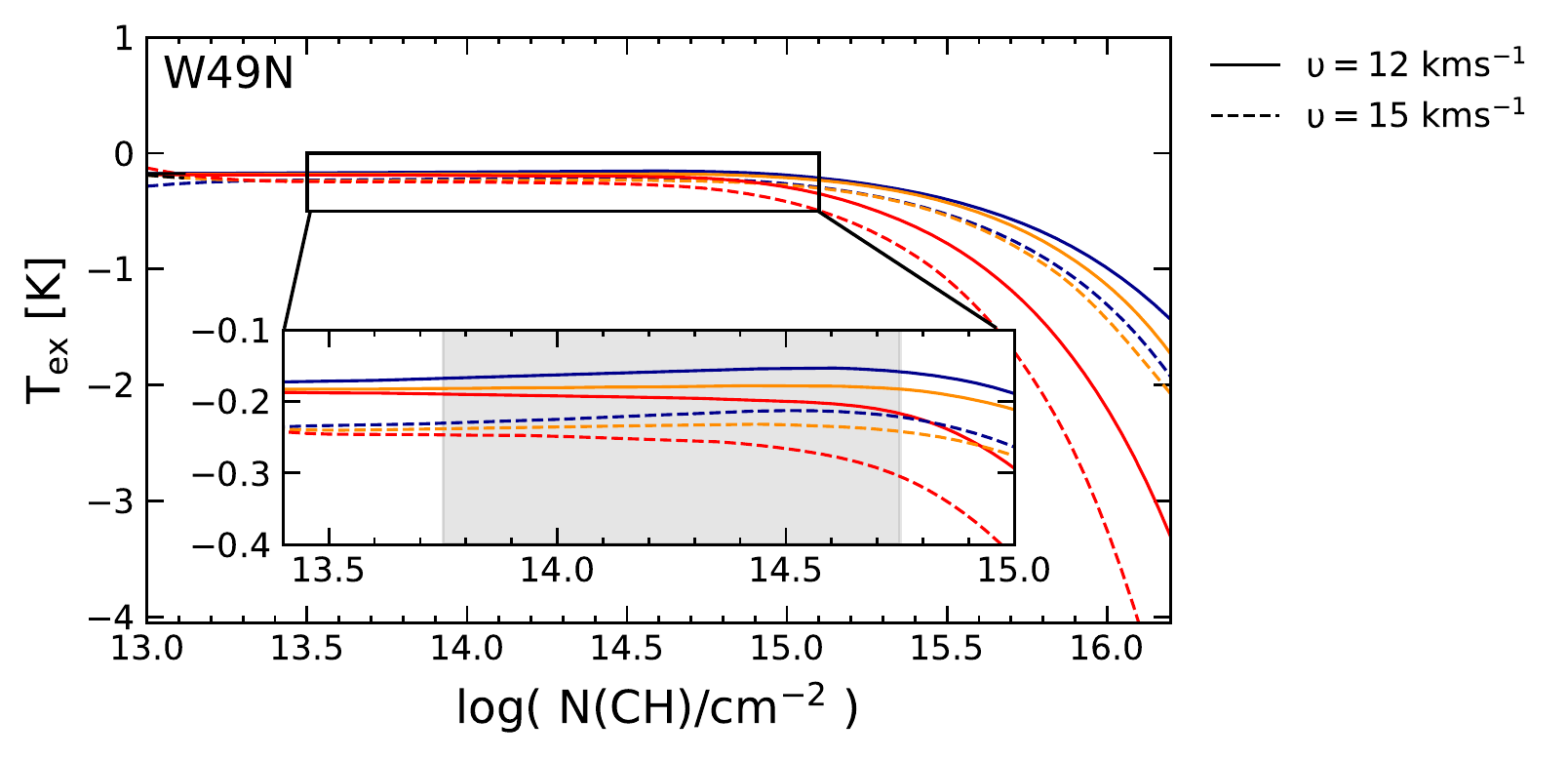}\quad
    \includegraphics[width=0.5\textwidth]{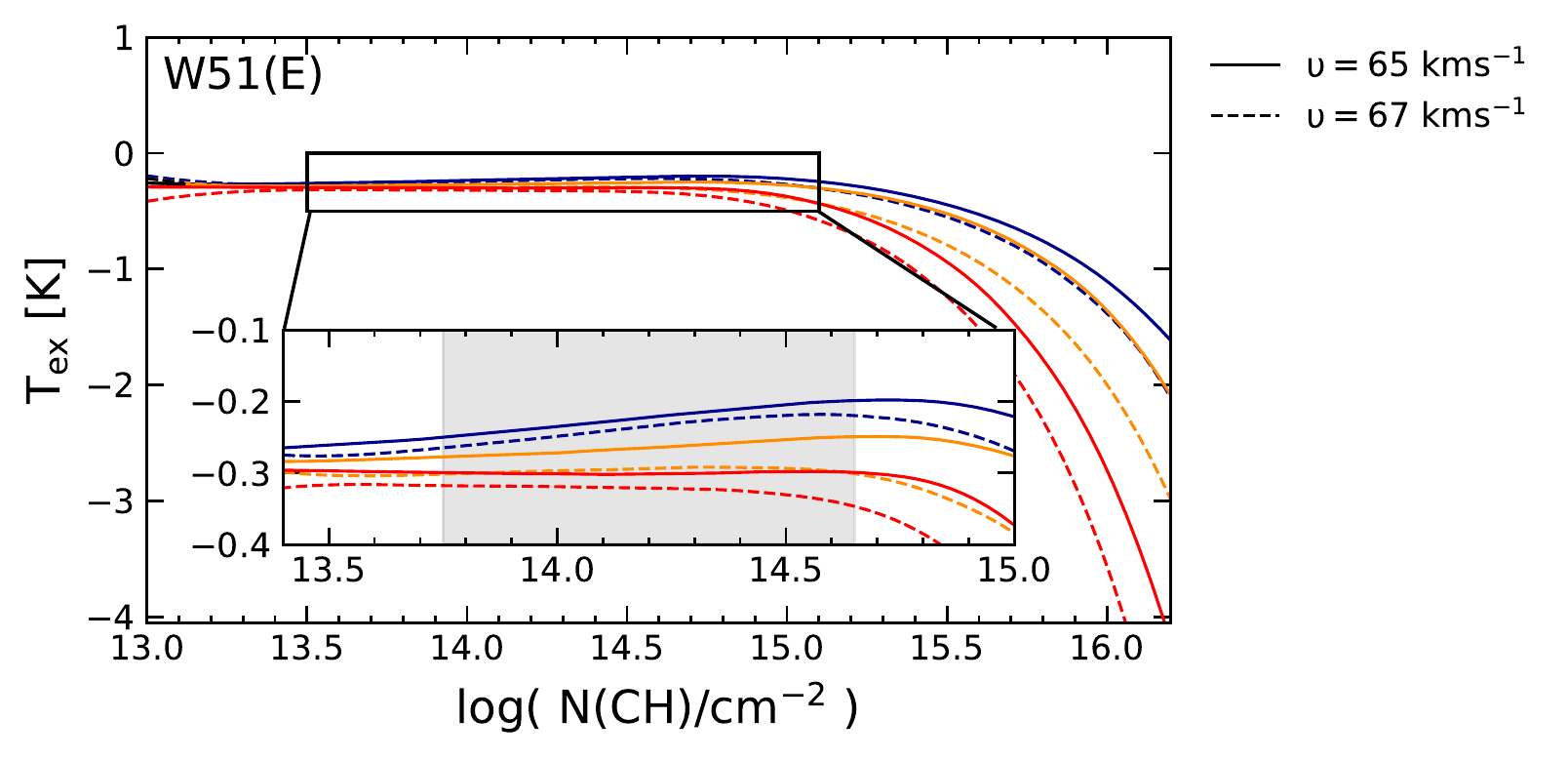}
    \caption{Clockwise from top-left: Modelled excitation temperatures of the 3.264~GHz (dark blue), 3.349~GHz (dark orange), and 3.335~GHz (red) ground state transitions of CH as a function of CH column densities for the physical conditions derived for the different velocity components toward Sgr~B2~(M), G34.26+0.15, W51~E, and W49~(N). The differing line styles are used to indicate results for the different velocity components as labelled. The inset panels expand on the $T_{\rm ex}$ values for CH column densities between $3\times10^{13}$ and 10$^{15}$~cm$^{-2}$ and the grey shaded regions highlight the column density intervals relevant to the velocity components in each respective source.}
    \label{fig:excitation_temp}
\end{figure*}

Using the physical conditions derived by the models discussed in Sect.~\ref{subsec:phy_conds}, we investigate the excitation conditions of each of the ground state HFS transitions of CH. In Fig.~\ref{fig:excitation_temp} we display the modelled excitation temperatures of the three HFS lines of CH corresponding to the physical conditions which characterise the cloud components toward each of the sources presented in this study. The low but negative excitation temperatures inferred from the models reflect level inversion in all three of the CH ground state HFS lines. Unsurprisingly, the degree of inversion (which corresponds to the strength of the excitation temperature) is only marginally higher in the lower satellite line and is the weakest in the main line. Additionally, by comparing the LOS and envelope velocity components observed toward Sgr~B2~(M), we find the LOS components to show weaker excitation temperatures which in turn suggests weaker masing effects, consistent with our observations. 
Our results are consistent with the theoretical range of excitation temperatures determined by \citet{Bertojo1976} and \citet{Elitzur1977}, between $-1.1$~K and $-0.5$~K, which is likely fortuitous, given that these studies did not have realistic collisional rate coefficients at their disposal. They are also in the range of values derived by \citet{Magnani1992} from observational data of outflow regions and with those derived by \citet{Genzel1979} toward SFRs, within the large quoted uncertainties. In a recent study, \citet{Dailey2020} derived a range of excitation temperatures by comparing emission in the CH 3.335~GHz radio HFS line emission with existing CH 4300~\AA\ optical absorption data for 16 lines of sight toward nearby stars. These authors note significant variations in the 3.335~GHz line's excitation temperature depending on the LOS and also emphasise that the assumption that $|T_{\rm ex}|$ is greater than the background radiation temperature is not always true, which is consistent with the results found in this work. Additionally across all the models discussed above, over the entire parameter space, all three HFS lines in the CH ground state are optically thin ($-0.01 <\tau <0$) such that the total amplification of the lines is always small in comparison to that in other widely observed interstellar maser lines from, for example, the OH and H$_{2}$O molecules, whose observationally determined brightness temperatures, which are $10^{12}$~K or higher, require much greater `maser gains', that is, amplification factors, $|\tau|> 20$ \citep[see for example, ][]{ReidMoran1981, Elitzur1982}.

\begin{figure}
\centering 
    \includegraphics[width=0.5\textwidth]{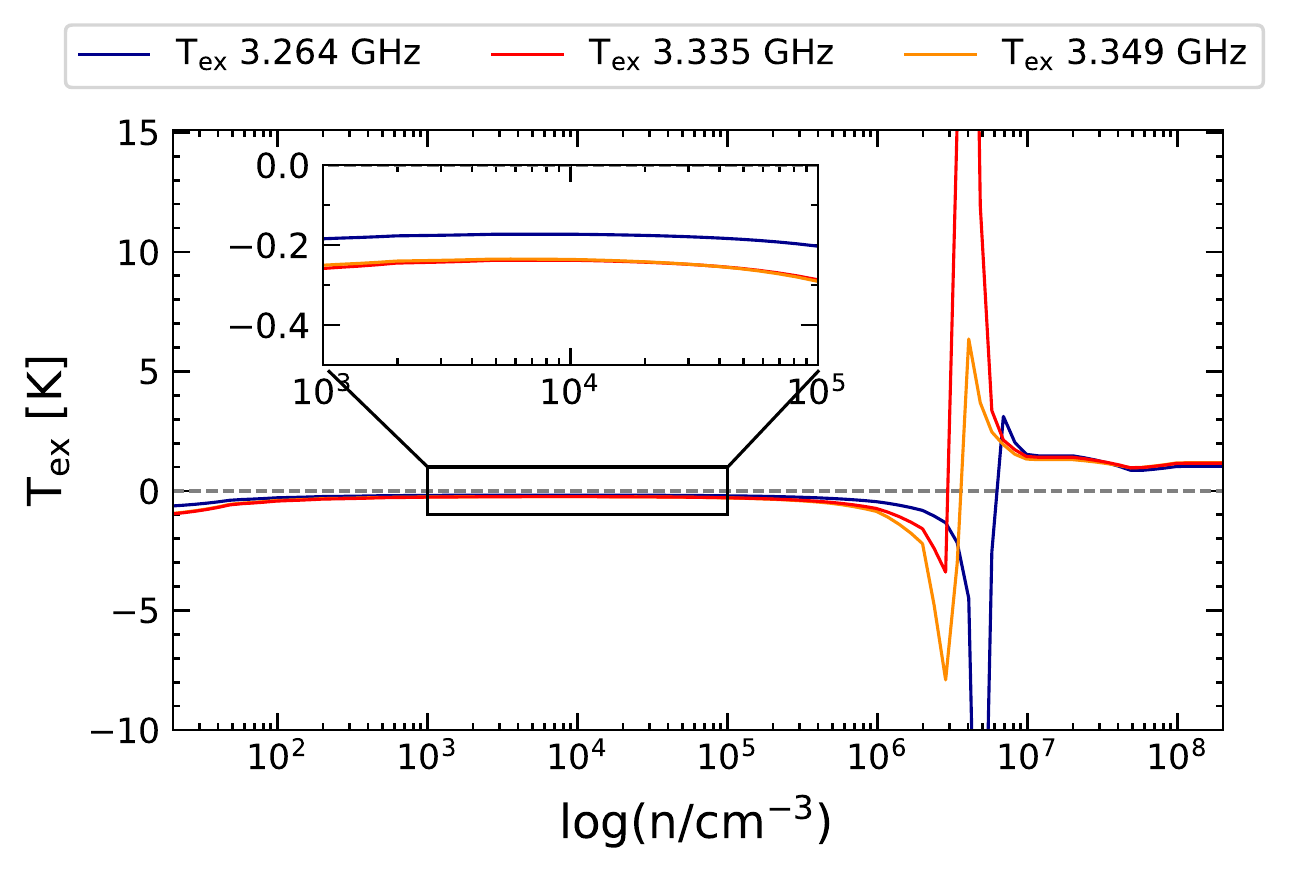}
    \caption{Variation in the excitation temperatures of the 3.264~GHz (dark blue), 3.349~GHz (dark orange), and 3.335~GHz (red) ground state transitions of CH as a function of gas densities at $T_{\rm kin} =75~$K, $N(\text{CH})=3\times10^{14}~$cm$^{-2}$ and $\Delta\upsilon = 6$~km~s$^{-1}$. The dashed grey line marks 0~K while the inset panel expands between $n$=$1\times10^{3}$ and $1\times10^{5}~$cm$^{-3}$.}
    \label{fig:tex_vs_dens}
\end{figure}

Moreover, the HFS lines in question have very low critical densities of ${\sim\!3}$ and 10~cm$^{-3}$ for collisions with molecular and atomic hydrogen at $T_{\rm kin}=50~$K, owing to their rather low Einstein A coefficients ${\sim 10^{-10}}~$s$^{-1}$. This in turn implies that the ground state transitions of CH are easily thermalised. In Fig.~\ref{fig:tex_vs_dens}, we analyse the trend between the excitation temperatures of the CH ground state HFS transitions and the gas densities for an envelope cloud model with a fixed gas temperature, CH column density and line width of 75~K, $3\times10^{14}~$cm$^{-2}$, and 6~km~s$^{-1}$, respectively. However, we find that the radio lines are not as easily thermalised and do not approach thermalisation in the range of gas densities modelled, here. 
The model reproduces negative excitation temperatures (or level inversion) up to gas densities of ${\!\sim\!4\times10^{6}}~$cm$^{-3}$, beyond which the excitation temperature changes drastically, displaying a sharp decrease followed by a rapid increase while flipping its sign. This density broadly corresponds to the critical density of the HFS lines of the $N,J=1,3/2\rightarrow1,1/2$ rotational transition near 532 and 537~GHz. Therefore, at gas densities above the critical density of the 532/537~GHz transitions de-excitation by collisional processes compete with radiative ones. This decreases the amount of radiative decay in these lines and in particular that of the $F^{\prime}$--$F^{\prime\prime}$=$2^{+}$-$0^{-}$ line,  which is partially responsible for the observed over population in the $F=0^{-}$ level of the CH ground state and explains the flip and subsequent increase in the excitation temperatures of the CH ground state HFS lines at higher densities. This once again emphasises the role played by the FIR line overlap in the anomalous excitation of the CH ground state HFS level. It also demonstrates that using the critical density, be it in its commonly used definition or also including radiative excitation, as a criterion for the thermalisation only holds for a two level system, but not for the more involved energy levels of CH. Moreover, the fact that the CH lines are not thermalised may paint a more complex picture in which the actual densities required for themalisation are much higher as discussed in Sect.~\ref{subsec:nlte_models}. Collisional excitation by electrons, which is not taken into account in our models, might be responsible for bringing these sub-thermally excited lines into thermalisation. 

\section{Conclusions}\label{sec:conclusions}
In this study we present the first interferometric observations of the 3.3~GHz (9~cm) HFS lines of the ground state $\Lambda$-doublet of CH. Despite being widely observed since 1973, the anomalous excitation of these lines, which  ubiquitously show weak maser action, has remained puzzling. While pumping cycles involving collisional excitation to the first rotational level following radiative decay back to the ground state could qualitatively explain the observed level inversions, they could not account for the relative intensities observed between the HFS lines and in particular the observed enhancement in the lower satellite line. We investigate the physical and excitation conditions of these lines using non-LTE radiative transfer analysis using recently derived collisional rate coefficients. The enhancement in the intensities of the lower satellite line are accounted for via the inclusion of pumping effects resulting from the radiative trapping caused by the overlap of FIR lines. By exploiting the synergy between the ground state and the $N, J =2,3/2\rightarrow1,1/2$ transitions of CH at 2006~GHz (149~$\mu$m) observed using upGREAT/SOFIA, we obtain additional constraints on the column densities of the different velocity components along each sight line. Modelling only the most prominent features detected in the spectra of all three HFS lines toward each sight line, which typically corresponds to velocity components associated with the envelope of the observed SFRs, we derive gas densities characteristic of translucent clouds ($~5\times10^{2}$--$10^{4}~$cm$^{-3}$) and gas temperatures between {$\sim\!50$ and 125}~K. The gas temperatures derived from the models are consistent with the average kinetic temperature (68$\pm$15)~K that \citet{Rachford2002} determined from Far Ultraviolet Spectroscopic Explorer (FUSE) observations of H$_2$ for a sample of translucent clouds. Moreover, the elevated temperatures traced by CH can be produced via the equilibrium heating by small grains in quiescent gas \citep{Wolfire2003} or alternatively, may result from this molecule's formation via non-thermal processes such as the deposition of turbulent energy into the gas. 
This makes CH a good tracer of the warm intermediary PDR, or TDR layers alongside CH$_{2}$, which we recently studied \citep{jacob2021}. The calculations presented here greatly depend upon the accuracy of the collisional rate coefficients used, while the modelling could certainly be improved by more realistic accounting for the structure of the target SFRs, like their thermal and density structure and the abundance gradient of CH in the envelopes in which it resides. Ultimately, a complete model of the CH ground state would require the inclusion of collisions with all relevant collision partners.

\subsection*{Outlook}
The constraints on the excitation conditions of the ground state HFS lines of CH provided by modelling such as ours will aid to firmly establish this molecule's widely and relatively easily observable set of HFS lines at 9~cm as a powerful radio-wavelength probe of the diffuse and translucent ISM, alongside the 21~cm H{\small I}, the 18~cm OH, and the 6~cm H$_{2}$CO lines. Observing the radio lines is a much more economical means of studying this radical, as observations of its sub-mm and FIR lines, while possible with GREAT, will always be expensive in terms of the limited observing time available with SOFIA, which is the only platform. As to the 3.3~GHz radio lines, in addition to single dishes, for northern sources (with declination ${>\!40^{\circ}}$), the VLA is well suited for observing them interferometrically. In addition, the southern sky is becoming available with the MeerKAT array in South Africa whose 64 antennas are currently being equipped with S-band receivers that cover frequencies from 1.7 to 3.5~GHz developed and built by the Max Planck Institute for Radio Astronomy \citep{Kramer2016}. 

An open question that remains is, how critical a role the sub-mm/FIR lines play in constraining the models of the 3.3~GHz ground state lines. On face value, it appears obvious that the value for the CH column densities they provide is essential for constraining the number of free parameters of a radiative transfer code such as MOLPOP-CEP,
which considers the source to be divided into zones of varying optical depths or column densities. Nevertheless, for sources that lack observations of the FIR lines of CH and therefore the accurate column density measurements they deliver, one can carry out an analysis by invoking models that describe the abundance distribution of the studied molecule across the modelled region. The abundance is often described by using a radial power-law distribution. Several studies have previously employed such an approach when carrying out non-LTE radiative transfer calculations as presented by \citet{Wyrowski2012, Wyrowski2016}, who use the Monte Carlo code RATRAN to model SOFIA/GREAT observations of FIR NH$_{3}$ absorption lines to trace infall in SFRs. Whether such an approach can lead to a meaningful  characterisation of the CH excitation and abundance based on the 3.3\,GHz radio lines alone is currently being explored.

\begin{acknowledgements}
We thank Sarantos Marinakis for providing us with collisional rate coefficients and discussions on the same. We are grateful to the referee, Harvey Liszt, for a careful review of the manuscript of this article and his insightful comments which have helped to improve its clarity. A.M.J. would like to thank Hans Nguyen for useful discussions at different stages of the data reduction process. The authors would like to express their gratitude to the developers of the many C++ and Python libraries, made available as open-source software; in particular this research has made use of the NumPy \citep{numpy}, SciPy \citep{scipy} and matplotlib \citep{matplotlib} packages. The National Radio Astronomy Observatory is a facility of
the National Science Foundation operated under cooperative agreement by Associated Universities, Inc. SOFIA Science Mission Operations is jointly operated by the Universities Space Research Association, Inc., under NASA contract NAS2-97001, and the Deutsches SOFIA Institut under DLR contract 50 OK 0901 and 50 OK 1301 to the University of Stuttgart. upGREAT is financed by resources from the participating institutes, and by the Deutsche Forschungsgemeinschaft (DFG) within the grant for the Collaborative Research Centre 956, as well as by the Federal Ministry of Economics and Energy (BMWI) via the German Space Agency (DLR) under Grants 50 OK 1102, 50 OK 1103 and 50 OK 1104.  
\end{acknowledgements}   
%
\bibliographystyle{aa} 
\bibliography{ref} 

\newcommand{\noop}[1]{}
\begin{thebibliography}{105}
\expandafter\ifx\csname natexlab\endcsname\relax\def\natexlab#1{#1}\fi

\bibitem[{{Alves} \& {Homeier}(2003)}]{Alves2003}
{Alves}, J. \& {Homeier}, N. 2003, \apjl, 589, L45

\bibitem[{{Asensio Ramos} \& {Elitzur}(2018)}]{Asensio2018}
{Asensio Ramos}, A. \& {Elitzur}, M. 2018, \aap, 616, A131

\bibitem[{{Astropy Collaboration} {et~al.}(2018){Astropy Collaboration},
  {Price-Whelan}, {Sip{\H{o}}cz}, {G{\"u}nther}, {Lim}, {Crawford}, {Conseil},
  {Shupe}, {Craig}, {Dencheva}, {Ginsburg}, {VanderPlas}, {Bradley},
  {P{\'e}rez-Su{\'a}rez}, {de Val-Borro}, {Aldcroft}, {Cruz}, {Robitaille},
  {Tollerud}, {Ardelean}, {Babej}, {Bach}, {Bachetti}, {Bakanov}, {Bamford},
  {Barentsen}, {Barmby}, {Baumbach}, {Berry}, {Biscani}, {Boquien}, {Bostroem},
  {Bouma}, {Brammer}, {Bray}, {Breytenbach}, {Buddelmeijer}, {Burke},
  {Calderone}, {Cano Rodr{\'\i}guez}, {Cara}, {Cardoso}, {Cheedella}, {Copin},
  {Corrales}, {Crichton}, {D'Avella}, {Deil}, {Depagne}, {Dietrich}, {Donath},
  {Droettboom}, {Earl}, {Erben}, {Fabbro}, {Ferreira}, {Finethy}, {Fox},
  {Garrison}, {Gibbons}, {Goldstein}, {Gommers}, {Greco}, {Greenfield},
  {Groener}, {Grollier}, {Hagen}, {Hirst}, {Homeier}, {Horton}, {Hosseinzadeh},
  {Hu}, {Hunkeler}, {Ivezi{\'c}}, {Jain}, {Jenness}, {Kanarek}, {Kendrew},
  {Kern}, {Kerzendorf}, {Khvalko}, {King}, {Kirkby}, {Kulkarni}, {Kumar},
  {Lee}, {Lenz}, {Littlefair}, {Ma}, {Macleod}, {Mastropietro}, {McCully},
  {Montagnac}, {Morris}, {Mueller}, {Mumford}, {Muna}, {Murphy}, {Nelson},
  {Nguyen}, {Ninan}, {N{\"o}the}, {Ogaz}, {Oh}, {Parejko}, {Parley}, {Pascual},
  {Patil}, {Patil}, {Plunkett}, {Prochaska}, {Rastogi}, {Reddy Janga},
  {Sabater}, {Sakurikar}, {Seifert}, {Sherbert}, {Sherwood-Taylor}, {Shih},
  {Sick}, {Silbiger}, {Singanamalla}, {Singer}, {Sladen}, {Sooley},
  {Sornarajah}, {Streicher}, {Teuben}, {Thomas}, {Tremblay}, {Turner},
  {Terr{\'o}n}, {van Kerkwijk}, {de la Vega}, {Watkins}, {Weaver}, {Whitmore},
  {Woillez}, {Zabalza}, \& {Astropy Contributors}}]{Astropy2018}
{Astropy Collaboration}, {Price-Whelan}, A.~M., {Sip{\H{o}}cz}, B.~M., {et~al.}
  2018, \aj, 156, 123

\bibitem[{{Astropy Collaboration} {et~al.}(2013){Astropy Collaboration},
  {Robitaille}, {Tollerud}, {Greenfield}, {Droettboom}, {Bray}, {Aldcroft},
  {Davis}, {Ginsburg}, {Price-Whelan}, {Kerzendorf}, {Conley}, {Crighton},
  {Barbary}, {Muna}, {Ferguson}, {Grollier}, {Parikh}, {Nair}, {Unther},
  {Deil}, {Woillez}, {Conseil}, {Kramer}, {Turner}, {Singer}, {Fox}, {Weaver},
  {Zabalza}, {Edwards}, {Azalee Bostroem}, {Burke}, {Casey}, {Crawford},
  {Dencheva}, {Ely}, {Jenness}, {Labrie}, {Lim}, {Pierfederici}, {Pontzen},
  {Ptak}, {Refsdal}, {Servillat}, \& {Streicher}}]{Astropy2013}
{Astropy Collaboration}, {Robitaille}, T.~P., {Tollerud}, E.~J., {et~al.} 2013,
  \aap, 558, A33

\bibitem[{{Bertojo} {et~al.}(1976){Bertojo}, {Cheung}, \&
  {Townes}}]{Bertojo1976}
{Bertojo}, M., {Cheung}, A.~C., \& {Townes}, C.~H. 1976, \apj, 208, 914

\bibitem[{{Bonfand} {et~al.}(2017){Bonfand}, {Belloche}, {Menten}, {Garrod}, \&
  {M{\"u}ller}}]{Bonfand2017}
{Bonfand}, M., {Belloche}, A., {Menten}, K.~M., {Garrod}, R.~T., \&
  {M{\"u}ller}, H.~S.~P. 2017, \aap, 604, A60

\bibitem[{{Bouloy} {et~al.}(1984){Bouloy}, {Nguyen-Q-Rieu}, \&
  {Field}}]{Bouloy1984}
{Bouloy}, D., {Nguyen-Q-Rieu}, \& {Field}, D. 1984, \aap, 130, 380

\bibitem[{{Bouloy} \& {Omont}(1977)}]{Bouloy1977}
{Bouloy}, D. \& {Omont}, A. 1977, \aap, 61, 405

\bibitem[{{Bouloy} \& {Omont}(1979)}]{Bouloy1979}
{Bouloy}, D. \& {Omont}, A. 1979, \aaps, 38, 101

\bibitem[{{Bujarrabal} {et~al.}(1984){Bujarrabal}, {Salinas}, \&
  {Gonzalo}}]{Bujarrabal1984}
{Bujarrabal}, V., {Salinas}, F., \& {Gonzalo}, I. 1984, \apj, 285, 312

\bibitem[{{Burton}(1970)}]{Burton1970}
{Burton}, W.~B. 1970, \aaps, 2, 291

\bibitem[{{Carpenter} \& {Sanders}(1998)}]{Carpenter1998}
{Carpenter}, J.~M. \& {Sanders}, D.~B. 1998, \aj, 116, 1856

\bibitem[{{Dagdigian}(2018{\natexlab{a}})}]{Dagdigian2018c2h}
{Dagdigian}, P.~J. 2018{\natexlab{a}}, \mnras, 479, 3227

\bibitem[{{Dagdigian}(2018{\natexlab{b}})}]{Dagdigian2018}
{Dagdigian}, P.~J. 2018{\natexlab{b}}, \mnras, 475, 5480

\bibitem[{{Dailey} {et~al.}(2020){Dailey}, {Smith}, {Magnani}, {Andersson}, \&
  {Reach}}]{Dailey2020}
{Dailey}, E.~M., {Smith}, A.~J., {Magnani}, L., {Andersson}, B.~G., \& {Reach},
  W.~T. 2020, \mnras, 495, 510

\bibitem[{{De Pree} {et~al.}(1997){De Pree}, {Mehringer}, \&
  {Goss}}]{depree1997}
{De Pree}, C.~G., {Mehringer}, D.~M., \& {Goss}, W.~M. 1997, \apj, 482, 307

\bibitem[{{De Pree} {et~al.}(2015){De Pree}, {Peters}, {Mac Low}, {Wilner},
  {Goss}, {Galv{\'a}n-Madrid}, {Keto}, {Klessen}, \& {Monsrud}}]{Depree2015}
{De Pree}, C.~G., {Peters}, T., {Mac Low}, M.~M., {et~al.} 2015, \apj, 815, 123

\bibitem[{{De Pree} {et~al.}(2020){De Pree}, {Wilner}, {Kristensen},
  {Galv{\'a}n-Madrid}, {Goss}, {Klessen}, {Mac Low}, {Peters}, {Robinson},
  {Sloman}, \& {Rao}}]{depree2020}
{De Pree}, C.~G., {Wilner}, D.~J., {Kristensen}, L.~E., {et~al.} 2020, \aj,
  160, 234

\bibitem[{{Dixon} \& {Field}(1979{\natexlab{a}})}]{Dixon1979}
{Dixon}, R.~N. \& {Field}, D. 1979{\natexlab{a}}, \mnras, 189, 583

\bibitem[{{Dixon} \& {Field}(1979{\natexlab{b}})}]{Dixon1979inelastic}
{Dixon}, R.~N. \& {Field}, D. 1979{\natexlab{b}}, Proceedings of the Royal
  Society of London Series A, 368, 99

\bibitem[{{Dreher} {et~al.}(1984){Dreher}, {Johnston}, {Welch}, \&
  {Walker}}]{Dreher1984}
{Dreher}, J.~W., {Johnston}, K.~J., {Welch}, W.~J., \& {Walker}, R.~C. 1984,
  \apj, 283, 632

\bibitem[{{Dunham}(1937)}]{dunham1937interstellar}
{Dunham}, T., J. 1937, \pasp, 49, 26

\bibitem[{{Ebenstein} \& {Muenter}(1984)}]{Ebenstein1984}
{Ebenstein}, W.~L. \& {Muenter}, J.~S. 1984, \jcp, 80, 3989

\bibitem[{{Elitzur}(1977)}]{Elitzur1977}
{Elitzur}, M. 1977, \apj, 218, 677

\bibitem[{{Elitzur}(1982)}]{Elitzur1982}
{Elitzur}, M. 1982, Reviews of Modern Physics, 54, 1225

\bibitem[{{Elitzur} \& {Asensio Ramos}(2006)}]{Elitzur2006}
{Elitzur}, M. \& {Asensio Ramos}, A. 2006, \mnras, 365, 779

\bibitem[{{Etxaluze} {et~al.}(2013){Etxaluze}, {Goicoechea}, {Cernicharo},
  {Polehampton}, {Noriega-Crespo}, {Molinari}, {Swinyard}, {Wu}, \&
  {Bally}}]{Etxaluze2013}
{Etxaluze}, M., {Goicoechea}, J.~R., {Cernicharo}, J., {et~al.} 2013, \aap,
  556, A137

\bibitem[{{Falgarone} {et~al.}(2005){Falgarone}, {Verstraete}, {Pineau Des
  For{\^e}ts}, \& {Hily-Blant}}]{Falgarone2005}
{Falgarone}, E., {Verstraete}, L., {Pineau Des For{\^e}ts}, G., \&
  {Hily-Blant}, P. 2005, \aap, 433, 997

\bibitem[{{Faure} {et~al.}(2007){Faure}, {Varambhia}, {Stoecklin}, \&
  {Tennyson}}]{Faure2007}
{Faure}, A., {Varambhia}, H.~N., {Stoecklin}, T., \& {Tennyson}, J. 2007,
  \mnras, 382, 840

\bibitem[{{Federman}(1982)}]{federman1982measurements}
{Federman}, S.~R. 1982, \apj, 257, 125

\bibitem[{{Felenbok} \& {Roueff}(1996)}]{Felenbok1996}
{Felenbok}, P. \& {Roueff}, E. 1996, \apjl, 465, L57

\bibitem[{{Genzel} {et~al.}(1979){Genzel}, {Downes}, {Pauls}, {Wilson}, \&
  {Bieging}}]{Genzel1979}
{Genzel}, R., {Downes}, D., {Pauls}, T., {Wilson}, T.~L., \& {Bieging}, J.
  1979, \aap, 73, 253

\bibitem[{{Gerin} {et~al.}(2010){Gerin}, {de Luca}, {Goicoechea}, {Herbst},
  {Falgarone}, {Godard}, {Bell}, {Coutens}, {Ka{\'z}mierczak}, {Sonnentrucker},
  {Black}, {Neufeld}, {Phillips}, {Pearson}, {Rimmer}, {Hassel}, {Lis},
  {Vastel}, {Boulanger}, {Cernicharo}, {Dartois}, {Encrenaz}, {Giesen},
  {Goldsmith}, {Gupta}, {Gry}, {Hennebelle}, {Hily-Blant}, {Joblin},
  {Ko{\l}os}, {Kre{\l}owski}, {Mart{\'\i}n-Pintado}, {Monje}, {Mookerjea},
  {Perault}, {Persson}, {Plume}, {Salez}, {Schmidt}, {Stutzki}, {Teyssier},
  {Yu}, {Contursi}, {Menten}, {Geballe}, {Schlemmer}, {Morris}, {Hatch},
  {Imram}, {Ward}, {Caux}, {G{\"u}sten}, {Klein}, {Roelfsema}, {Dieleman},
  {Schieder}, {Honingh}, \& {Zmuidzinas}}]{gerin2010interstellar}
{Gerin}, M., {de Luca}, M., {Goicoechea}, J.~R., {et~al.} 2010, \aap, 521, L16

\bibitem[{{Ginsburg} {et~al.}(2017){Ginsburg}, {Goddi}, {Kruijssen}, {Bally},
  {Smith}, {Galv{\'a}n-Madrid}, {Mills}, {Wang}, {Dale}, {Darling},
  {Rosolowsky}, {Loughnane}, {Testi}, \& {Bastian}}]{Ginsburg2017}
{Ginsburg}, A., {Goddi}, C., {Kruijssen}, J.~M.~D., {et~al.} 2017, \apj, 842,
  92

\bibitem[{{Godard} {et~al.}(2012){Godard}, {Falgarone}, {Gerin}, {Lis}, {De
  Luca}, {Black}, {Goicoechea}, {Cernicharo}, {Neufeld}, {Menten}, \&
  {Emprechtinger}}]{Godard2012}
{Godard}, B., {Falgarone}, E., {Gerin}, M., {et~al.} 2012, \aap, 540, A87

\bibitem[{{Godard} {et~al.}(2009){Godard}, {Falgarone}, \& {Pineau Des
  For{\^e}ts}}]{Godard2009}
{Godard}, B., {Falgarone}, E., \& {Pineau Des For{\^e}ts}, G. 2009, \aap, 495,
  847

\bibitem[{{Godard} {et~al.}(2014){Godard}, {Falgarone}, \& {Pineau des
  For{\^e}ts}}]{godard2014chemical}
{Godard}, B., {Falgarone}, E., \& {Pineau des For{\^e}ts}, G. 2014, \aap, 570,
  A27

\bibitem[{{Goldsmith} \& {Kauffmann}(2017)}]{Goldsmith2017}
{Goldsmith}, P.~F. \& {Kauffmann}, J. 2017, \apj, 841, 25

\bibitem[{{Goldsmith} {et~al.}(2010){Goldsmith}, {Velusamy}, {Li}, \&
  {Langer}}]{Goldsmith2010}
{Goldsmith}, P.~F., {Velusamy}, T., {Li}, D., \& {Langer}, W.~D. 2010, \apj,
  715, 1370

\bibitem[{{Goss} \& {Field}(1968)}]{Goss1968}
{Goss}, W.~M. \& {Field}, G.~B. 1968, \apj, 151, 177

\bibitem[{{Greaves} \& {Williams}(1994)}]{Greaves1994}
{Greaves}, J.~S. \& {Williams}, P.~G. 1994, \aap, 290, 259

\bibitem[{Harris {et~al.}(2020)Harris, Millman, van~der Walt, Gommers,
  Virtanen, Cournapeau, Wieser, Taylor, Berg, Smith, Kern, Picus, Hoyer, van
  Kerkwijk, Brett, Haldane, Fernández~del Río, Wiebe, Peterson,
  Gérard-Marchant, Sheppard, Reddy, Weckesser, Abbasi, Gohlke, \&
  Oliphant}]{numpy}
Harris, C.~R., Millman, K.~J., van~der Walt, S.~J., {et~al.} 2020, Nature, 585,
  357–362

\bibitem[{{Heaton} {et~al.}(1989){Heaton}, {Little}, \& {Bishop}}]{Heaton1989}
{Heaton}, B.~D., {Little}, L.~T., \& {Bishop}, I.~S. 1989, \aap, 213, 148

\bibitem[{{Hierl} {et~al.}(1997){Hierl}, {Morris}, \&
  {Viggiano}}]{hierl1997rate}
{Hierl}, P.~M., {Morris}, R.~A., \& {Viggiano}, A.~A. 1997, \jcp, 106, 10145

\bibitem[{{Hunter}(2007)}]{matplotlib}
{Hunter}, J.~D. 2007, Computing in Science Engineering, 9, 90

\bibitem[{{Jacob} {et~al.}(2021){Jacob}, {Menten}, {Gong}, {Bergman}, {Tiwari},
  {Br{\"u}nken}, \& {Olofsson}}]{jacob2021}
{Jacob}, A.~M., {Menten}, K.~M., {Gong}, Y., {et~al.} 2021, \aap, 647, A42

\bibitem[{{Jacob} {et~al.}(2020){Jacob}, {Menten}, {Wiesemeyer}, {G{\"u}sten},
  {Wyrowski}, \& {Klein}}]{jacob2020}
{Jacob}, A.~M., {Menten}, K.~M., {Wiesemeyer}, H., {et~al.} 2020, \aap, 640,
  A125

\bibitem[{{Jacob} {et~al.}(2019){Jacob}, {Menten}, {Wiesemeyer}, {Lee},
  {G{\"u}sten}, \& {Dur{\'a}n}}]{jacob2019fingerprinting}
{Jacob}, A.~M., {Menten}, K.~M., {Wiesemeyer}, H., {et~al.} 2019, \aap, 632,
  A60

\bibitem[{Jones {et~al.}(2001)Jones, Oliphant, Peterson, {et~al.}}]{scipy}
Jones, E., Oliphant, T., Peterson, P., {et~al.} 2001, {SciPy}: Open source
  scientific tools for {Python}

\bibitem[{{K{\"o}nig} {et~al.}(2017){K{\"o}nig}, {Urquhart}, {Csengeri},
  {Leurini}, {Wyrowski}, {Giannetti}, {Wienen}, {Pillai}, {Kauffmann},
  {Menten}, \& {Schuller}}]{Konig2017}
{K{\"o}nig}, C., {Urquhart}, J.~S., {Csengeri}, T., {et~al.} 2017, \aap, 599,
  A139

\bibitem[{{Kramer} {et~al.}(2016){Kramer}, {Menten}, {Barr}, {Karuppusamy},
  {Kasemann}, {Klein}, {Ros}, {Wieching}, \& {Wucknitz}}]{Kramer2016}
{Kramer}, M., {Menten}, K., {Barr}, E.~D., {et~al.} 2016, in MeerKAT Science:
  On the Pathway to the SKA, 3

\bibitem[{{Kurayama} {et~al.}(2011){Kurayama}, {Nakagawa}, {Sawada-Satoh},
  {Sato}, {Honma}, {Sunada}, {Hirota}, \& {Imai}}]{Kurayama2011}
{Kurayama}, T., {Nakagawa}, A., {Sawada-Satoh}, S., {et~al.} 2011, \pasj, 63,
  513

\bibitem[{{Lang} \& {Wilson}(1978)}]{Lang1978}
{Lang}, K.~R. \& {Wilson}, R.~F. 1978, \apj, 224, 125

\bibitem[{{Liszt} \& {Lucas}(1996)}]{Liszt1996}
{Liszt}, H. \& {Lucas}, R. 1996, \aap, 314, 917

\bibitem[{{Little} {et~al.}(1979){Little}, {MacDonald}, {Riley}, \&
  {Matheson}}]{Little1979}
{Little}, L.~T., {MacDonald}, G.~H., {Riley}, P.~W., \& {Matheson}, D.~N. 1979,
  \mnras, 189, 539

\bibitem[{{Magnani} \& {Onello}(1993)}]{Magnani1993}
{Magnani}, L. \& {Onello}, J.~S. 1993, \apj, 408, 559

\bibitem[{{Magnani} {et~al.}(1992){Magnani}, {Sandell}, \&
  {Lada}}]{Magnani1992}
{Magnani}, L., {Sandell}, G., \& {Lada}, E.~A. 1992, \aaps, 93, 509

\bibitem[{{Marinakis} {et~al.}(2019){Marinakis}, {Kalugina}, {K{\l}os}, \&
  {Lique}}]{Marinakis2019}
{Marinakis}, S., {Kalugina}, Y., {K{\l}os}, J., \& {Lique}, F. 2019, \aap, 629,
  A130

\bibitem[{{Mathis} {et~al.}(1983){Mathis}, {Mezger}, \& {Panagia}}]{Mathis1983}
{Mathis}, J.~S., {Mezger}, P.~G., \& {Panagia}, N. 1983, \aap, 500, 259

\bibitem[{{Mattila}(1986)}]{Mattila1986}
{Mattila}, K. 1986, \aap, 160, 157

\bibitem[{{McKellar}(1940)}]{McKellar1940}
{McKellar}, A. 1940, \pasp, 52, 187

\bibitem[{{Mookerjea} {et~al.}(2007){Mookerjea}, {Casper}, {Mundy}, \&
  {Looney}}]{Mookerjea2007}
{Mookerjea}, B., {Casper}, E., {Mundy}, L.~G., \& {Looney}, L.~W. 2007, \apj,
  659, 447

\bibitem[{{Neill} {et~al.}(2014){Neill}, {Bergin}, {Lis}, {Schilke},
  {Crockett}, {Favre}, {Emprechtinger}, {Comito}, {Qin}, {Anderson},
  {Burkhardt}, {Chen}, {Harris}, {Lord}, {McGuire}, {McNeill}, {Monje},
  {Phillips}, {Steber}, {Vasyunina}, \& {Yu}}]{Neill2014}
{Neill}, J.~L., {Bergin}, E.~A., {Lis}, D.~C., {et~al.} 2014, \apj, 789, 8

\bibitem[{{Phelps} \& {Dalby}(1966)}]{Phelps1966}
{Phelps}, D.~H. \& {Dalby}, F.~W. 1966, \prl, 16, 3

\bibitem[{{Qin} {et~al.}(2010){Qin}, {Schilke}, {Comito}, {M{\"o}ller},
  {Rolffs}, {M{\"u}ller}, {Belloche}, {Menten}, {Lis}, {Phillips}, {Bergin},
  {Bell}, {Crockett}, {Blake}, {Cabrit}, {Caux}, {Ceccarelli}, {Cernicharo},
  {Daniel}, {Dubernet}, {Emprechtinger}, {Encrenaz}, {Falgarone}, {Gerin},
  {Giesen}, {Goicoechea}, {Goldsmith}, {Gupta}, {Herbst}, {Joblin},
  {Johnstone}, {Langer}, {Lord}, {Maret}, {Martin}, {Melnick}, {Morris},
  {Murphy}, {Neufeld}, {Ossenkopf}, {Pagani}, {Pearson}, {P{\'e}rault},
  {Plume}, {Salez}, {Schlemmer}, {Stutzki}, {Trappe}, {van der Tak}, {Vastel},
  {Wang}, {Yorke}, {Yu}, {Zmuidzinas}, {Boogert}, {G{\"u}sten}, {Hartogh},
  {Honingh}, {Karpov}, {Kooi}, {Krieg}, {Schieder}, {Diez-Gonzalez},
  {Bachiller}, {Martin-Pintado}, {Baechtold}, {Olberg}, {Nordh}, {Gill}, \&
  {Chattopadhyay}}]{Qin2010}
{Qin}, S.~L., {Schilke}, P., {Comito}, C., {et~al.} 2010, \aap, 521, L14

\bibitem[{{Rachford} {et~al.}(2002){Rachford}, {Snow}, {Tumlinson}, {Shull},
  {Blair}, {Ferlet}, {Friedman}, {Gry}, {Jenkins}, {Morton}, {Savage},
  {Sonnentrucker}, {Vidal-Madjar}, {Welty}, \& {York}}]{Rachford2002}
{Rachford}, B.~L., {Snow}, T.~P., {Tumlinson}, J., {et~al.} 2002, \apj, 577,
  221

\bibitem[{{Reid} \& {Ho}(1985)}]{Reid1985}
{Reid}, M.~J. \& {Ho}, P.~T.~P. 1985, \apjl, 288, L17

\bibitem[{{Reid} {et~al.}(2019){Reid}, {Menten}, {Brunthaler}, {Zheng}, {Dame},
  {Xu}, {Li}, {Sakai}, {Wu}, {Immer}, {Zhang}, {Sanna}, {Moscadelli}, {Rygl},
  {Bartkiewicz}, {Hu}, {Quiroga-Nu{\~n}ez}, \& {van Langevelde}}]{Reid2019}
{Reid}, M.~J., {Menten}, K.~M., {Brunthaler}, A., {et~al.} 2019, \apj, 885, 131

\bibitem[{{Reid} \& {Moran}(1981)}]{ReidMoran1981}
{Reid}, M.~J. \& {Moran}, J.~M. 1981, \araa, 19, 231

\bibitem[{{Risacher} {et~al.}(2016){Risacher}, {G{\"u}sten}, {Stutzki},
  {H{\"u}bers}, {Bell}, {Buchbender}, {B{\"u}chel}, {Csengeri}, {Graf},
  {Heyminck}, {Higgins}, {Honingh}, {Jacobs}, {Klein}, {Okada}, {Parikka},
  {P{\"u}tz}, {Reyes}, {Ricken}, {Riquelme}, {Simon}, \&
  {Wiesemeyer}}]{risacher2016upgreat}
{Risacher}, C., {G{\"u}sten}, R., {Stutzki}, J., {et~al.} 2016, \aap, 595, A34

\bibitem[{{Robitaille} \& {Bressert}(2012)}]{AplPy2012}
{Robitaille}, T. \& {Bressert}, E. 2012, {APLpy: Astronomical Plotting Library
  in Python}

\bibitem[{{Rydbeck} {et~al.}(1973){Rydbeck}, {Elld{\'e}r}, \&
  {Irvine}}]{rydbeck1973}
{Rydbeck}, O.~E.~H., {Elld{\'e}r}, J., \& {Irvine}, W.~M. 1973, \nat, 246, 466

\bibitem[{{Rydbeck} {et~al.}(1976){Rydbeck}, {Kollberg}, {Hjalmarson}, {Sume},
  {Ellder}, \& {Irvine}}]{Rydbeck1976}
{Rydbeck}, O.~E.~H., {Kollberg}, E., {Hjalmarson}, A., {et~al.} 1976, \apjs,
  31, 333

\bibitem[{{Sakai} {et~al.}(2012){Sakai}, {Maezawa}, {Sakai}, {Menten}, \&
  {Yamamoto}}]{Sakai2012}
{Sakai}, N., {Maezawa}, H., {Sakai}, T., {Menten}, K.~M., \& {Yamamoto}, S.
  2012, \aap, 546, A103

\bibitem[{{S{\'a}nchez-Monge} {et~al.}(2017){S{\'a}nchez-Monge}, {Schilke},
  {Schmiedeke}, {Ginsburg}, {Cesaroni}, {Lis}, {Qin}, {M{\"u}ller}, {Bergin},
  {Comito}, \& {M{\"o}ller}}]{Sanchez2017}
{S{\'a}nchez-Monge}, {\'A}., {Schilke}, P., {Schmiedeke}, A., {et~al.} 2017,
  \aap, 604, A6

\bibitem[{{Sandell} {et~al.}(1988){Sandell}, {Magnani}, \&
  {Lada}}]{Sandell1988}
{Sandell}, G., {Magnani}, L., \& {Lada}, E.~A. 1988, \apj, 329, 920

\bibitem[{{Sato} {et~al.}(2000){Sato}, {Hasegawa}, {Whiteoak}, \&
  {Miyawaki}}]{Sato2000}
{Sato}, F., {Hasegawa}, T., {Whiteoak}, J.~B., \& {Miyawaki}, R. 2000, \apj,
  535, 857

\bibitem[{{Sato} {et~al.}(2010){Sato}, {Reid}, {Brunthaler}, \&
  {Menten}}]{sato2010}
{Sato}, M., {Reid}, M.~J., {Brunthaler}, A., \& {Menten}, K.~M. 2010, \apj,
  720, 1055

\bibitem[{{Schewe} {et~al.}(2015){Schewe}, {Ma}, {Vanhaecke}, {Wang},
  {K{\l}os}, {Alexander}, {van de Meerakker}, {Meijer}, {van der Avoird}, \&
  {Dagdigian}}]{Schewe2015}
{Schewe}, H.~C., {Ma}, Q., {Vanhaecke}, N., {et~al.} 2015, \jcp, 142, 204310

\bibitem[{{Schmiedeke} {et~al.}(2016){Schmiedeke}, {Schilke}, {M{\"o}ller},
  {S{\'a}nchez-Monge}, {Bergin}, {Comito}, {Csengeri}, {Lis}, {Molinari},
  {Qin}, \& {Rolffs}}]{Schmiedeke2016}
{Schmiedeke}, A., {Schilke}, P., {M{\"o}ller}, T., {et~al.} 2016, \aap, 588,
  A143

\bibitem[{{Schuller} {et~al.}(2009){Schuller}, {Menten}, {Contreras},
  {Wyrowski}, {Schilke}, {Bronfman}, {Henning}, {Walmsley}, {Beuther},
  {Bontemps}, {Cesaroni}, {Deharveng}, {Garay}, {Herpin}, {Lefloch}, {Linz},
  {Mardones}, {Minier}, {Molinari}, {Motte}, {Nyman}, {Reveret}, {Risacher},
  {Russeil}, {Schneider}, {Testi}, {Troost}, {Vasyunina}, {Wienen}, {Zavagno},
  {Kovacs}, {Kreysa}, {Siringo}, \& {Wei{\ss}}}]{Schuller2009}
{Schuller}, F., {Menten}, K.~M., {Contreras}, Y., {et~al.} 2009, \aap, 504, 415

\bibitem[{{Serabyn} {et~al.}(1993){Serabyn}, {Guesten}, \&
  {Schulz}}]{Serabyn1993}
{Serabyn}, E., {Guesten}, R., \& {Schulz}, A. 1993, \apj, 413, 571

\bibitem[{{Sheffer} {et~al.}(2008){Sheffer}, {Rogers}, {Federman}, {Abel},
  {Gredel}, {Lambert}, \& {Shaw}}]{sheffer2008}
{Sheffer}, Y., {Rogers}, M., {Federman}, S.~R., {et~al.} 2008, \apj, 687, 1075

\bibitem[{{Sievers} {et~al.}(1991){Sievers}, {Mezger}, {Bordeon}, {Kreysa},
  {Haslam}, \& {Lemke}}]{Sievers1991}
{Sievers}, A.~W., {Mezger}, P.~G., {Bordeon}, M.~A., {et~al.} 1991, \aap, 251,
  231

\bibitem[{{Snow} \& {McCall}(2006)}]{Snow2006}
{Snow}, T.~P. \& {McCall}, B.~J. 2006, \araa, 44, 367

\bibitem[{{Stacey} {et~al.}(1987){Stacey}, {Lugten}, \& {Genzel}}]{Stacey1987}
{Stacey}, G.~J., {Lugten}, J.~B., \& {Genzel}, R. 1987, \apj, 313, 859

\bibitem[{{Suutarinen} {et~al.}(2011){Suutarinen}, {Geppert}, {Harju},
  {Heikkil{\"a}}, {Hotzel}, {Juvela}, {Millar}, {Walsh}, \&
  {Wouterloot}}]{Suutarinen2011}
{Suutarinen}, A., {Geppert}, W.~D., {Harju}, J., {et~al.} 2011, \aap, 531, A121

\bibitem[{{Swings} \& {Rosenfeld}(1937)}]{swings1937}
{Swings}, P. \& {Rosenfeld}, L. 1937, \apj, 86, 483

\bibitem[{{Thiel} {et~al.}(2019){Thiel}, {Belloche}, {Menten}, {Giannetti},
  {Wiesemeyer}, {Winkel}, {Gratier}, {M{\"u}ller}, {Colombo}, \&
  {Garrod}}]{Thiel2019A}
{Thiel}, V., {Belloche}, A., {Menten}, K.~M., {et~al.} 2019, \aap, 623, A68

\bibitem[{{Truppe} {et~al.}(2014){Truppe}, {Hendricks}, {Hinds}, \&
  {Tarbutt}}]{Truppe2014}
{Truppe}, S., {Hendricks}, R.~J., {Hinds}, E.~A., \& {Tarbutt}, M.~R. 2014,
  \apj, 780, 71

\bibitem[{{Vastel} {et~al.}(2001){Vastel}, {Spaans}, {Ceccarelli}, {Tielens},
  \& {Caux}}]{Vastel2001}
{Vastel}, C., {Spaans}, M., {Ceccarelli}, C., {Tielens}, A.~G.~G.~M., \&
  {Caux}, E. 2001, \aap, 376, 1064

\bibitem[{{Watson}(2001)}]{watson2001assignment}
{Watson}, J. K.~G. 2001, \apj, 555, 472

\bibitem[{{Welch} {et~al.}(1987){Welch}, {Dreher}, {Jackson}, {Terebey}, \&
  {Vogel}}]{Welch1987}
{Welch}, W.~J., {Dreher}, J.~W., {Jackson}, J.~M., {Terebey}, S., \& {Vogel},
  S.~N. 1987, Science, 238, 1550

\bibitem[{{Weselak}(2019)}]{Weselak2019}
{Weselak}, T. 2019, \aap, 625, A55

\bibitem[{{Whiteoak} {et~al.}(1980){Whiteoak}, {Gardner}, \&
  {Hoglund}}]{Whiteoak1980}
{Whiteoak}, J.~B., {Gardner}, F.~F., \& {Hoglund}, B. 1980, \mnras, 190, 17P

\bibitem[{{Wiesemeyer} {et~al.}(2018){Wiesemeyer}, {G{\"u}sten}, {Menten},
  {Dur{\'a}n}, {Csengeri}, {Jacob}, {Simon}, {Stutzki}, \&
  {Wyrowski}}]{wiesemeyer2018unveiling}
{Wiesemeyer}, H., {G{\"u}sten}, R., {Menten}, K.~M., {et~al.} 2018, \aap, 612,
  A37

\bibitem[{{Winkel} {et~al.}(2017){Winkel}, {Wiesemeyer}, {Menten}, {Sato},
  {Brunthaler}, {Wyrowski}, {Neufeld}, {Gerin}, \&
  {Indriolo}}]{winkel2017hydrogen}
{Winkel}, B., {Wiesemeyer}, H., {Menten}, K.~M., {et~al.} 2017, \aap, 600, A2

\bibitem[{{Wolfire} {et~al.}(2003){Wolfire}, {McKee}, {Hollenbach}, \&
  {Tielens}}]{Wolfire2003}
{Wolfire}, M.~G., {McKee}, C.~F., {Hollenbach}, D., \& {Tielens}, A.~G.~G.~M.
  2003, \apj, 587, 278

\bibitem[{{Wyrowski} {et~al.}(2016){Wyrowski}, {G{\"u}sten}, {Menten},
  {Wiesemeyer}, {Csengeri}, {Heyminck}, {Klein}, {K{\"o}nig}, \&
  {Urquhart}}]{Wyrowski2016}
{Wyrowski}, F., {G{\"u}sten}, R., {Menten}, K.~M., {et~al.} 2016, \aap, 585,
  A149

\bibitem[{{Wyrowski} {et~al.}(2012){Wyrowski}, {G{\"u}sten}, {Menten},
  {Wiesemeyer}, \& {Klein}}]{Wyrowski2012}
{Wyrowski}, F., {G{\"u}sten}, R., {Menten}, K.~M., {Wiesemeyer}, H., \&
  {Klein}, B. 2012, \aap, 542, L15

\bibitem[{{Xu} \& {Li}(2016)}]{Xu2016}
{Xu}, D. \& {Li}, D. 2016, \apj, 833, 90

\bibitem[{{Young} {et~al.}(2012){Young}, {Becklin}, {Marcum}, {Roellig}, {De
  Buizer}, {Herter}, {G{\"u}sten}, {Dunham}, {Temi}, {Andersson}, {Backman},
  {Burgdorf}, {Caroff}, {Casey}, {Davidson}, {Erickson}, {Gehrz}, {Harper},
  {Harvey}, {Helton}, {Horner}, {Howard}, {Klein}, {Krabbe}, {McLean}, {Meyer},
  {Miles}, {Morris}, {Reach}, {Rho}, {Richter}, {Roeser}, {Sandell}, {Sankrit},
  {Savage}, {Smith}, {Shuping}, {Vacca}, {Vaillancourt}, {Wolf}, \&
  {Zinnecker}}]{young2012early}
{Young}, E.~T., {Becklin}, E.~E., {Marcum}, P.~M., {et~al.} 2012, \apjl, 749,
  L17

\bibitem[{{Zhang} {et~al.}(2013){Zhang}, {Reid}, {Menten}, {Zheng},
  {Brunthaler}, {Dame}, \& {Xu}}]{zhang2013parallaxes}
{Zhang}, B., {Reid}, M.~J., {Menten}, K.~M., {et~al.} 2013, \apj, 775, 79

\bibitem[{{Zhang} {et~al.}(2009){Zhang}, {Zheng}, {Reid}, {Menten}, {Xu},
  {Moscadelli}, \& {Brunthaler}}]{zhang2009trigonometric}
{Zhang}, B., {Zheng}, X.~W., {Reid}, M.~J., {et~al.} 2009, \apj, 693, 419

\bibitem[{{Zuckerman} \& {Turner}(1975)}]{Zuckerman1975}
{Zuckerman}, B. \& {Turner}, B.~E. 1975, \apj, 197, 123

\end{thebibliography}
%

\begin{appendix}
\section{Impact of using different collisional rate coefficients}\label{sec:collisional_rate}
In this appendix we illustrate the results obtained when using collisional rate coefficients computed directly for collisions of CH with para-H$_{2}$. The excitation temperatures are computed over a range of CH column densities but for a fixed gas temperature and line width of 75~K and 6~km~s$^{-1}$ for two sets of gas densities at 5$\times10^{2}$ and 5$\times10^{3}$~cm$^{-3}$. We see that for both cases the MOLPOP-CEP models reproduce positive values for the excitation temperatures of all three of the CH ground state HFS lines until CH column densities of ${\sim \!10^{14}}~$cm$^{-2}$. Beyond this value the lower satellite shows a rapid increase in the excitation temperature following which it decreases to negative values thereby displaying level inversion in this line. 
\begin{figure}
     
    \includegraphics[width=0.49\textwidth]{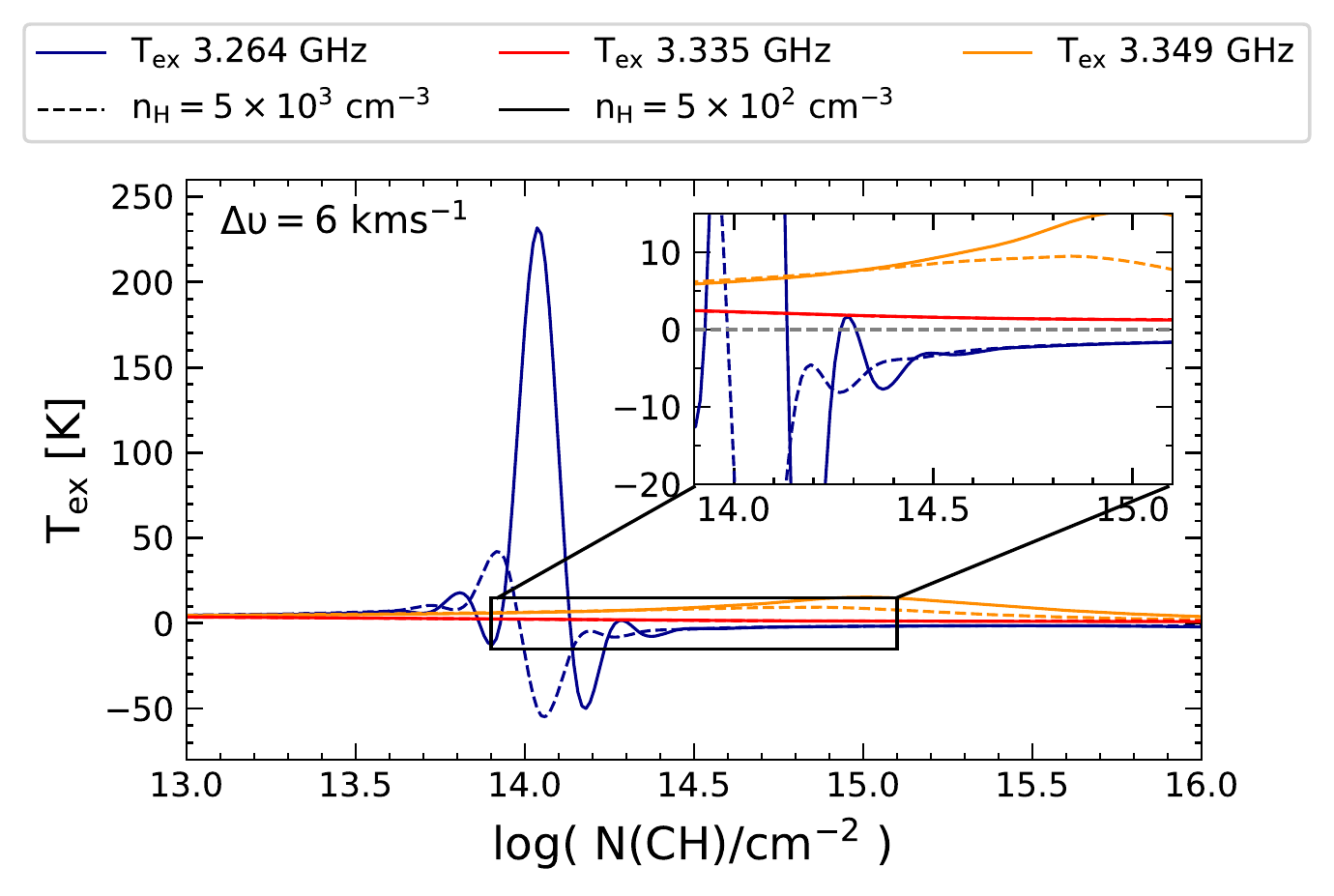}
    \caption{MOLPOP-CEP non-LTE radiative transfer model of the ground state HFS lines of CH using a combination of collisional rate coefficients for collisions of CH with ortho-H$_{2}$, para-H$_{2}$, and atomic hydrogen all computed by \citet{Dagdigian2018}. The solid and dashed blue, red and dark orange curves represent the excitation temperatures of the 3.264~GHz, 3.335~GHz and 3.349~GHz CH lines, as a function of $N({\rm CH})$ for $n= 5\times10^{2}$, and $5\times10^{3}~$cm$^{-3}$, respectively for models with $T_{\rm kin} = 75~$K and $\Delta\upsilon = 6~$km~s$^{-1}$. The inset panel expands on the $T_{\rm ex}$ values for CH column densities between $10^{14}$--10$^{15}$~cm$^{-2}$.}
    \label{fig:other_models}
\end{figure}
\end{appendix}
\end{document}